%% file: ds4.tex
\documentstyle[seceq,epsf]{ptptex}

\title{
Dilaton Stabilization in (A)dS Spacetime with Compactified Dimensions
}

\author{
Kunihito {\sc Uzawa}\footnote{E-mail: uzawa@phys.h.kyoto-u.ac.jp} 
}

\inst{
Graduate School of Human and Environmental Studies,
Kyoto University, Kyoto 606-8501, Japan}

\recdate{March 7, 2003
}

\abst{

\input{abstract}

}

\begin{document}

\maketitle


\input{sec1}
\input{sec2}
\input{sec3}
\input{sec4}
\input{sec5}

\appendix
\input{appenA}
\input{appenB}
\input{appenC}

\input{appenD}
\input{ref}

\end{document}

%% file: abstract.tex
We investigate dilaton stabilization in a higher-dimensional theory.
The background geometry is based on an eleven-dimensional 
Kaluza-Klein/supergravity model, which is assumed to be a product of
four-dimensional de Sitter ($dS_4$)  spacetime and a seven  
sphere ($S^7$). The dilaton potential has a local minimum resulting from  
contributions of the cosmological constant, the curvature of the
internal spacetime and quantum effects of the background
scalar, vector, spinor, and tensor fields. The dilaton settles down to the
 local minimum, and the scale of
 the extra dimensions eventually become time independent. 
Our four-dimensional universe evolves from $dS_4$ into $AdS_4$
after stabilization of the extra dimension.

%% file: sec1.tex

\section{Introduction}
	\label{sec:introduction}

In a unified description of gravity and gauge
interactions, the extra dimension plays a very important role in
providing a consistent quantum theory of all known interactions.
One of the ideas for this unification is that four-dimensional
gauge interactions arise from isometries of the higher-dimensional
gravity. In the general strategy usually employed, it is considered
 that obvious difference between the usual four dimensions and
 the extra dimension could result from a process
of spontaneous breakdown of the vacuum symmetry, which is often called
``spontaneous compactification''\cite{cr1} of the extra dimensions.
This process has been actively investigated in
connection with supergravity.\cite{cr}${}^{,}$\cite{en}  

 However, the modern version of the unification scheme is much different
from the conventional Kaluza-Klein approach.\cite{ka}${}^{,}$\cite{kl} 
Such a modification is motivated by
several factors. First, we cannot naturally understand the hierarchy
involving the electroweak scale and the fundamental scale in particle
physics. Second, it has been shown that the dimensionally reduced
bare action directly obtained from higher-dimensional gravity cannot
give rise to a stable compactification. A new compactification mechanism, 
recently proposed by Randall and Sundrum (RS),\cite{ra}  has been studied
 intensively. 
This is different from the conventional Kaluza-Klein picture, in
which the 
gauge and matter field are assumed to be confined on the ``brane,''
 while the gravity and moduli exist in the ``bulk.'' 
However, because the RS 
prescription at present cannot be applied directly to more than
 five dimensions, the Kaluza-Klein  description is still considered
 useful. 
In fact, Kaluza-Klein supergravity theories\cite{du} 
are regarded as low energy effective version of M-theory.
The extra dimensions in the Kaluza-Klein description are assumed to be 
microscopic, with a size much smaller than the scale of four dimensions.
Constraints on the smallness of the extra dimensions have been obtained
with the accelerator experiments. Specifically, 
the upper bound on this has found in this way is well known to be 
approximately $10^{-16}$ cm. The extra dimensions must be then 
not only  small but also almost static. 
The gauge coupling constant is in general related to
 the ratio of the higher-dimensional gravitational constant and the 
Kaluza-Klein mass.\cite{malc} 
The Kaluza-Klein mass depends  on the size of 
extra dimensions. Hence the time dependence of the extra dimensions 
induces  that of the gauge coupling constants.
At present, the time variation of gauge coupling constants is constrained
by the measurement of the quasar absorption line,
\cite{le}${}^{-}$\cite{mu} the cosmic 
microwave background (CMB), \cite{kap}${}^{,}$\cite{av} 
and primordial nucleosynthesis,\cite{be} according to 
which their change is at least very small after
 nucleosynthesis.\cite{ko} We thus see that
 the stabilization of the dilaton is a crucial issue. 

The process of spontaneous compactification generates a degree of
  freedom of the dilaton field,
  which is characterized by the scale of the extra dimensions in lower
dimensions. 
This implies that the dynamics of the extra dimensions are governed
 by the dilaton potential. Then, in order to stabilize the extra
dimensions, the dilaton potential must have a minimum.   
Unfortunately, in a system with $D$-dimensional ($D>4$)
 gravity (or gravity plus a scalar) 
 and a cosmological constant, classically the dilaton potential 
 has no local minimum in a naive compactification,
 such as the product of the four-dimensional spacetime and a 
$(D-4)$ sphere or torus.

Recently, Carroll et al. proposed a dilaton stabilization
mechanism in a classical system\cite{ca}. 
The essence of their idea is to consider a 
 combination of background matter fields.
The dilaton potential in their model consists of the background
 energy-momentum
tensor, which contains the curvature of the extra dimensions,
a cosmological constant and a Yang-Mills field, which
 is wrapped around the extra dimensions.   
Positive curvature effects are contributed to the attractive force 
for the dilaton,  
while the cosmological constant and the Yang-Mills field strength form
 a repulsive force,\footnote{The magnetic strength is sort of
 $U(1)$ fiber bundle. The strength of the repulsive force
 for the extra dimension increases 
because the force arising from the magnetic strength is
 proportional to the flux density. } so that the   
 dilaton is stabilized by the balance of
these forces. Torii and Shiromizu also studied stabilization of
the Freund-Rubin type of compactification with a $p$-form gauge field
 strength.\cite{to}

In this paper, taking into account quantum effects on a compactified
spacetime, we consider the dynamics of the extra dimension in the
 four-dimensional de Sitter spacetime ($dS_4$). 
First, we consider the system consisting 
of scalar, vector, spinor and tensor fields
 and a cosmological constant in eleven dimensions.
The eleven-dimensional space is assumed to break up into a
 four-dimensional
 de Sitter space and a curved compact seven-dimensional manifold. 
The energy-momentum tensor in this model consists of the matter
and its quantum effects. 
This quantum correction is often called the 
Casimir effect.\cite{ap}${}^{,}$\cite{ch}  
The quantum effective potential associated with the matter fields 
is similar power to   
the part of dilaton potential arising from the 
curvature and the cosmological constant.\cite{can}  
However, their signs are opposite, so that 
a local minimum is created by their  combination.  
Taking account of this result, we investigate the stabilization in the
bosonic part of the eleven-dimensional supergravity model.
The energy-momentum tensor possesses not only classical fields but also  
the contribution of the 1-loop quantum corrections in graviton,
 because the physics associate with the extra 
dimension in the Kaluza-Klein theories is comparable to or not too
 much larger than the Planck length. 
Therefore, the quantum effects of moduli are presumably of some
 importance, 
and these effects are able to contain Einstein-Hilbert action,  
which is assumed to be the low-energy effective action of the M-theory.
\cite{m}

To evaluate the functional determinant in this calculation of the 1-loop 
quantum effect, we use the generalized 
zeta function, which is the sum of the operator eigenvalues.
Because the de Sitter spacetime has a unique Euclidean section which is an
$S^4$, we can assume that the grand state of this system is the
product  $S^4\times S^7$ of a four-dimensional sphere
 and a seven-dimensional sphere.  We will see that 
the zeta function regularization method is useful.
Finally, we show that the Casimir effect
 produces a potential minimum and moduli field is stabilized. 

 The plan of this paper is as follows. A brief description of 
 $D$-dimensional Kaluza-Klein gravity is given in $\S$\ref{sec:classical}.
In $\S$\ref{sec:quantum}, 
we calculate the effective action due to one loop of the
scalar, vector, spinor and tensor fields in eleven dimensions,
 which is a product 
of four-dimensional de Sitter spacetime and $S^7$. 
We use these results in $\S$\ref{sec:sugra} to carry out an 
analysis of the stability of the dilaton in eleven-dimensional
 supergravity model.            
Section \ref{sec:conclusion} contains our conclusions.

%% file: sec2.tex

\section{Extra dimension classical dynamics in a 
$D$-dimensional model}
	\label{sec:classical}

\subsection{Gravity plus scalar field system}
        \label{subsec: gravity plus scalar}
First, as a simple example, we review the classical
 dynamics of a dilaton field in a $D (D>4)$-dimensional model.
The background geometry is given by the product space of
 four-dimensional spacetime and the $(D-4)$-dimensional sphere $S^d$.
The four-dimensional metric is not specified here.  
We obtain a massive mode of the dilaton
 after the background spacetime is compactified on $S^d$. 
This model is simple but important for studying the dynamics of
 dilaton fields. 

We consider the following $D$-dimensional Einstein-Hilbert action with a
cosmological constant:  
%
\begin{equation}
I_{\rm EH}=\frac{1}{2\bar{\kappa}^2} \int d^Dx
\sqrt{-\bar{g}}(\bar{R}-2\bar{\Lambda})\,.
\label{eq;2-1}
\end{equation}
Here, $ \bar{\kappa} $ is a positive constant, $\bar{R}$ is the
$D$-dimensional Ricci scalar, and $\bar{\Lambda}$ is the cosmological 
constant. We consider the case of a positive cosmological
constant, $\Lambda>0$.
The line element is now assumed to take the form
%
\begin{equation}
ds^2=\hat{g}_{\mu\nu}dx^{\mu}dx^{\nu}+b^2\Omega^{(d)}_{ij}dx^idx^j\,,
\label{eq;2-2}
\end{equation}
where $\hat{g}_{\mu\nu}$ is a four-dimensional metric depending
 only on the four-dimensional coordinates $\{x^{\mu}\}
 \,\,(\mu=0, 1, 2, 3)$,  and
$\Omega^{(d)}_{ij}dx^idx^j$ is a line element of unit $d$-dimensional
sphere. The dilaton field, $b$, which is a so-called ``radion'', 
 characterizes the scale of the extra dimension and is assumed to depend
 only on the four-dimensional coordinate $x^{\mu}$.  
Substituting the metric (\ref{eq;2-2}) into the action (\ref{eq;2-1}),
 we obtain a four-dimensional action  
%
\begin{eqnarray}
 I_{\rm EH}&=&\frac{1}{2\kappa^2}\int d^4x\sqrt{-\hat{g}}
	\left(\frac{b}{b_{0}}\right)^d\left[ \hat{R} + 
        Kd(d-1)\hat{g}^{\mu\nu}(\partial_{\mu}\ln b)(\partial_{\nu}\ln b)
         \right.\nonumber\\
       & &\left.+ d(d-1)b^{-2} -2\bar{\Lambda}\right]\,,
\label{eq;2-3}
\end{eqnarray}
where $\kappa$ is a positive constant defined by 
$\kappa^2=\bar{\kappa}^2/(2^d\,b_{0}^d\,\pi)$, $K\,(K>0)$ is the curvature
parameter of $S^d$, and $\hat{R}$ is the Ricci scalar 
of the four-dimensional 
metric tensor $\hat{g}_{\mu\nu}$. Because we 
consider the case in which the scale $b$ of the compactification 
is a time-dependent 
variable, this action is different from the ordinary four-dimensional
 Einstein-Hilbert action. 
For this reason, we carry out the following conformal transformation
 in order to obtain a useful expression:
%
\begin{equation}
 \hat{g}_{\mu\nu} = (b/b_{0})^{-d}g_{\mu\nu}\,.
\label{eq;2-4}
\end{equation}
Note that if we take $b_0$ to be the initial value of $b$, then the
conformal factor is initially unity. 
After the conformal transformation, which changes the 
$D$-dimensional line element to 
%
\begin{equation}
\hat{g}_{MN}dx^Mdx^N=\left(\frac{b}{b_0}\right)^{-d}
g_{\mu\nu}dx^{\mu}dx^{\nu}+ b^2\Omega_{ij}^{(d)}dx^idx^j\,,
           \label{eq;2-m}
\end{equation}
we obtain
%
\begin{equation}
 I_{\rm EH}=\int d^4x\sqrt{-g}\left[\frac{1}{2\kappa^2}R
	-\frac{1}{2}g^{\mu\nu}\partial_{\mu}\sigma\partial_{\nu}\sigma 
	- U_{0}(\sigma)\right]\,,
           \label{eq;2-5}
\end{equation}
where $R$ is the Ricci scalar of the four-dimensional metric tensor 
$g_{\mu\nu}$, the field $\sigma$ is defined by 
%
\begin{equation}
 \sigma  =  \sigma_{0}\ln\left(\frac{b}{b_{0}}\right)\,, 
	 \hspace{1cm}
 \sigma_{0}  =  \sqrt{\frac{d(d+2)}{2\kappa^2}}\,,
         \label{eq;sigma}
\end{equation}
and the potential $U_{0}(\sigma)$ of $\sigma$ is 
%
\begin{equation}
 U_{0}(\sigma) = \frac{\bar{\Lambda}}{\kappa^2}\,{\rm e}^{-d\sigma /\sigma_{0}}
 	-\frac{Kd(d-1)}{2\kappa^2b_{0}^2}\,{\rm e}^{-(d+2)\sigma /\sigma_{0}}\,.
          \label{eq;2-6}
\end{equation}
Unfortunately, it can be easily
confirmed that the dilaton field $\sigma$ [i.e. $b(t)$] potential 
$U_{0}(\sigma)$ does {\it not} possess a local minimum.  This implies that
there is no stable compactification on $S^{d}$ if the model is
modified no further.

\subsection{Gravity, 
the  cosmological constant and the gauge field system}
Next, we consider the contribution of the gauge field, in addition to 
the dilaton field and the cosmological constant. 
This system has been investigated by Carroll et al..\cite{ca} 
Although they considered the 
six-dimensional spacetime, motivated by the work of 
Cremmer and Scherk\cite{cr2} 
and Sundrum,\cite{sun} 
 we consider the general $D$-dimensional case.
The ansatz here for the line element is given by (\ref{eq;2-m}).
The starting point of this model is the 
Einstein-Hilbert action with the matter term
%
\begin{equation}
I_{\rm EH}=\frac{1}{2\bar{\kappa}^2} \int d^Dx
\sqrt{-\bar{g}}(\bar{R}-2\Lambda)+{\cal L}_{\rm m}\,,
\label{eq;2-7}
\end{equation}
where ${\cal L}_{\rm m}$ is the matter Lagrangian density. 
Using the line element (\ref{eq;2-2}) and the conformal transformation
(\ref{eq;2-4}), 
we obtain the four-dimensional action as
%
\begin{equation}
 I_{\rm EH}=\int d^4x\sqrt{-g}\left[\frac{1}{2\kappa^2}R
	-\frac{1}{2}g^{\mu\nu}\partial_{\mu}\sigma\partial_{\nu}\sigma 
	- V(\sigma)\right]\,,
\label{eq;2-8}
\end{equation}
where the dilaton potential $V(\sigma)$ is written  
%
\begin{equation}
V(\sigma)=\frac{1}{\kappa^2}\left\{-\frac{Kd(d-1)}{b_0^2}
          \,{\rm e}^{-(d+2)\sigma/\sigma_0}
          +{\rm e}^{-d\sigma/\sigma_0}
          \left(\Lambda-{\cal L}_m\right)\right\}\,.
\label{eq;2-9}
\end{equation}
We introduce the electromagnetic fields $F_{\mu\nu}$ as matter fields.
The explicit form of $F_{\mu\nu}$ is the Freund-Rubin type, \cite{fr}
$F_{ij}=F_{ji}=\left(\epsilon_{ij}F_0\right)\sqrt{\Omega}$,
 where $i$ and $j$ run over the extra dimensions. The moduli potential is 
%
\begin{equation}
V(\sigma)=\frac{1}{\kappa^2}\left\{-\frac{Kd(d-1)}{b_0^2}\,
          {\rm e}^{-(d+2)\sigma/\sigma_0}
          +{\rm e}^{-d\sigma/\sigma_0}\Lambda
          +f_0^2\,{\rm e}^{-(d+4)\sigma/\sigma_0}\right\}\,,
\label{eq;2-10}
\end{equation}
where $f_0^2=F_0^2\Omega/2$.  We search for values of the parameters 
$\Lambda$ and 
$f_0$ such that there exists a local minimum
 of the potential $V(\sigma)$. 
Following the approach of Carroll et al., \cite{ca}
 we choose the parameters as 
%
\begin{equation}
\frac{Kd(d-1)}{4\kappa^2b_0^2}=\frac{\Lambda}{\kappa^2}
=\frac{f_0^2}{\kappa^2}\equiv A\,,
\label{eq;2-11}
\end{equation}
where $A$ is constant. The dilaton potential $V(\sigma)$ 
has a local minimum at $b=0$.  
 The dilaton acquires a mass 
$m_{\sigma}=d^2V/db^2|_{b=b_0}\propto 1/b_0^2$, 
and the energy scale at
the maximum of the potential is $V(\kappa^2\sigma)\simeq A$.
As remarked above, 
the dynamics of the extra dimension are to introduce an additional massive
dilaton field into the four-dimensional theory. 
Although the extra dimension is stabilized by the contribution of the 
background matter contents (i.e. the repulsive force of the 
magnetic flux and the cosmological constant, and the attractive force of
 the curvature of the extra dimension), 
the state $b=0$ is not a global minimum, so that there is the possibility
 that the dilaton will tunnel
through the potential barrier. 
Because the dilaton in that case runs away to infinity,
 there arises the problem of extra dimensions expanding to
infinity.\cite{bu}  

%% file: sec3.tex

\section{Quantum effect in the $D$-dimensional Kaluza-Klein model}
	\label{sec:quantum}
It is known that an attractive force exists between 
uncharged 
superconductivity plates as
 a consequence of the quantum mechanical vacuum 
fluctuations of the electromagnetic field. This is called the  
Casimir effect.\cite{cas}
Such a quantum correction is expected to arise in Kaluza-Klein theory, 
because a boundary condition is imposed on the quantum field in the
 direction of the extra dimension. This effect is very important in 
determining the dynamics of the extra dimension in Kaluza-Klein theories.
\cite{ap}${}^{,}$\cite{can}${}^{,}$\cite{maeda}
Here, we present a calculation method for the 1-loop effective potential
 generated by quantum
fluctuations of scalar, vector, Dirac spinor and tensor fields   
with $D$-dimensional spacetime given by the product of $n$-dimensional
de Sitter spacetime ($dS_n$) and the $d$-dimensional sphere ($S^d$). 
To carry out the 1-loop evaluation of the effective potential,
 the gauge-fixing and ghost terms are introduced. 
Then, we explain a convenient regularization based on 
the $\zeta$ function. 
As concrete examples of the method, calculations of the effective
potential in 
 the product space of four-dimensional Minkowski spacetime and the 
$d$-sphere ($M_4\times S^d$) and in $M_4\times S^n\times S^d$ 
have been presented previously by several 
authors.\cite{ki}${}^{-}$\cite{bi1}      
\subsection{Dimensional reduction}
As a simple example, we consider the action of the Einstein-Hilbert 
 and massless scalar field $\bar{\phi}$ in $D\,\,(=n+d)$ dimensions,
%
\begin{equation}
I[b,\phi]=I_{\rm EH}+I_{\rm S}\,,
\end{equation}
where $I_{\rm EH}$ is the Einstein-Hilbert action (\ref{eq;2-1}), and 
%
\begin{equation}
I_{\rm S}=-\frac{1}{2}\int d^{D}x\sqrt{-\bar{g}}\bar{g}^{MN}
 	\partial_{M}\bar{\phi}\partial_{N}\bar{\phi}
\end{equation}
is a massless Klein-Gordon field.
The background geometry is the $n$-dimensional de Sitter spacetime
 with $S^d$.
The $D$-dimensional line element here has the same form as 
 $(\ref{eq;2-4})$. 
The metric $\hat{g}_{\mu\nu}$ in (\ref{eq;2-4}) initially represents an
 $n$-dimensional de Sitter
 spacetime and depends only on the $n$-dimensional coordinates
 $\{x^{\mu}\}$.
The scalar field $\bar{\phi}$ is expanded as 
%
\begin{equation}
 \bar{\phi} = b_{0}^{-d/2}
	\sum_{l,m}\phi_{lm}(x^{\mu})Y^{(d)}_{lm}(x^i)\,,
\end{equation}
where the constant $b_{0}$ is the initial value of $b$.
In this 
expression, the $Y^{(d)}_{lm}$ are real harmonics 
on the $d$-sphere satisfying 
%
\begin{eqnarray}
 \frac{1}{\sqrt{\Omega^{(d)}}}\partial_{i}\left(
 	\sqrt{\Omega^{(d)}}\Omega^{(d)ij}\partial_{j}Y^{(d)}_{lm}\right)
	 + l(l+d-1)Y^{(d)}_{lm} 
 & = & 0\,,\\
 \int d^dx\sqrt{\Omega^{(d)}}Y^{(d)}_{lm}Y^{(d)}_{l'm'} 
 & = & \delta_{ll'}\delta_{mm'}\,,
\end{eqnarray}
and $\phi_{lm}(x^{\mu})$ is a real function depending only on 
the $n$-dimensional coordinates $\left\{x^{\mu}\right\}$. 
The massless scalar field action in this case is given by
%
\begin{equation}
 I_{\rm S} =  -\frac{1}{2}\sum_{l,m}\int d^nx\sqrt{-\hat{g}}\,
 	{\rm e}^{d\frac{\sigma}{\sigma_0}}\left[
 	\hat{g}^{\mu\nu}\partial_{\mu}\phi_{lm}
 	\partial_{\nu}\phi_{lm} 
 	+ \frac{l(l+d-1)}{b^2}\;\phi_{lm}^2
 	\right]\,.
   \label{eq;ms}
\end{equation}
The massless scalars in $D$ dimensions acquire mass through
 spontaneous compactification.
The massive scalar field action (\ref{eq;ms}) is quite different from the
usual $n$-dimensional form of that action, because the dilaton field
 couples to the 
kinetic term of the scalar field $\phi_{lm}$. Carrying out the
conformal transformation
%
\begin{equation}
\hat{g}_{\mu\nu}=\left(\frac{b}{b_0}\right)^{-2d/(n-2)}g_{\mu\nu}\,,
    \label{eq;cometric}
\end{equation}
where $g_{\mu\nu}$ is the metric on $dS_n$ and $b_0$ is chosen as the
initial value of $b$, we obtain 
%
\begin{equation}
 I_{\rm S} =  -\frac{1}{2}\sum_{l,m}\int d^nx\sqrt{-g}
 	\left[g^{\mu\nu}\partial_{\mu}\phi_{lm}
 	\partial_{\nu}\phi_{lm} 
 	+ M_{\phi}^2\;\phi_{lm}^2
 	\right]\,,
\end{equation}
where the mass $M^2_{\phi}$ of the $n$-dimensional scalar field
$\phi_{lm}$ is given by  
%
\begin{equation}
M_{\phi}^2=\frac{l(l+d-1)}{b_0^2}{\rm e}^{-2(d+n-2)\sigma/\{(n-2)\sigma_0\}}\,,
\end{equation}
$R$ is the Ricci scalar on the $dS_n$ background, and the
dilaton field $\sigma$ is also defined by Eq.\,(\ref{eq;sigma}). 


\subsection{The quantum correction}
     \label{subsec;quantum}
The calculation of the effective potential is carried out using the path
integral method.
The fields are split into a classical part, $\phi_{lm\;{\rm c}}$, 
and a quantum part, $\delta{\phi_{lm}}$. The action
is then expanded with respect to
 the quantum fields around an arbitrary classical
background field, up to second order, to
generate all 1-loop diagrams.
 
In the path integral approach to quantum field theory, the amplitude 
is given by the expression
%
\begin{equation}
Z=\int{\cal D}\phi_{lm}\;\exp{\left(iI_{\rm t}[b, \phi_{lm}]\right)}\,,
       \label{eq;subquantum-1}
\end{equation}
where ${\cal D}\phi_{lm}$ is the integration measure on the
 space of scalar fields, and $I_{\rm t}[b, \phi_{lm}]$ is the total action. 
The action can be expanded in the neighborhood of these classical
 background fields as
%
\begin{equation}
I_{\rm t}[b,\phi_{lm}]=I[b, \phi_{lm\;{\rm c}}]+I_{\rm q}[b, \delta{\phi_{lm}}]
+O\left((\delta \phi_{lm})^3\right)\,,
       \label{eq;subquantum-2}       
\end{equation}
where $\phi_{lm}=\phi_{lm\;{\rm c}}+\delta\phi_{lm}$. The action 
$I_{\rm q}[b,\phi_{lm}]$ is  
quadratic in $\delta\phi_{lm}$, and no terms linear in
 $\phi_{lm}$ appear, because such terms can be 
eliminated by using the classical equations of motion. 
In the 1-loop approximation, 
we ignore all terms of higher than quadratic order in the expansion of  
%
\begin{equation}
\ln Z=iI[b, \phi_{lm\:{\rm c}}]
   +\ln\left\{\int{\cal D}\delta\phi_{lm}\;
     \exp{\left(iI_{\rm q}[b, \delta\phi_{lm}]\right)}\right\}\,.
          \label{eq;3-2-2}
\end{equation}
We note that the integrals here are ill-defined, because the
 operators in this
equation are unbounded in the $dS_n$ spacetime
 with Lorentz
signature. We need to perform a Wick rotation in order to make 
 this expression well-defined. Doing so, we obtain
%
\begin{equation}
\ln Z=-I_{\rm E}[b, \phi_{lm\;{\rm c}}]
      +\ln\left\{\int{\cal D}\delta\phi_{lm}\;
       \exp{\left(-I_{\rm qE}[b, \delta\phi_{lm}]\right)
       }\right\}\,,
          \label{eq;3-2-3}
\end{equation}
where $I_{\rm E}$ is the Euclidean action, which is expressed as  
%
\begin{equation}
I_{\rm E}[b, \phi_{lm}]=\sum_{l,\;m}\int d^n x\sqrt{-g}
             \frac{1}{2}\left\{\partial_{\mu}\phi_{lm}
             \partial^{\mu}\phi_{lm}+M^2_{\phi}\phi^2_{lm}\right\}\,.
          \label{eq;3-2-4}
\end{equation}
Using the assumption $\phi_{lm}=\phi_{lm\;{\rm c}}+\delta\phi_{lm}$,
 we can integrate the kinetic term in the action by parts, obtaining
%
\begin{equation}
I_{\rm E}[b, \delta\phi_{lm}]=\frac{1}{2}\int d^n x\sqrt{-g}
             \delta\phi_{lm}\left\{-\Box_{(n)}
              +M^2_{\phi}\right\}\delta\phi_{lm}\,,
           \label{eq;3-2-5}
\end{equation}
where $\Box_{(n)}$ denotes the Laplacian in the $n$-dimensional de Sitter
 spacetime.  
 
The effective potential $V_{\rm eff}$ is defined through the relation  
%
\begin{eqnarray}
\exp\left(\int d^nx V_{\rm eff}\right)
   &=&\int {\cal D}\phi_{lm}
    \exp\left(-I_{\rm qE}[b, \delta\phi_{lm}]\right)\nonumber\\
   &=&\left\{\det\mu^{-2}\left(\Box_{(n)}-M^2_{\phi}\right)
     \right\}^{-\frac{1}{2}}\,,
  \label{eq;3-2-6}
\end{eqnarray}
where $I_{\rm qE}$ is the total Euclidean action and  $\mu$ is a 
normalization constant with dimension of mass
that comes from the Euclidean path integral.
Using Eqs. (\ref{eq;3-2-3}) and (\ref{eq;3-2-6}), 
we find that the 1-loop effective potential is 
%
\begin{equation}
V_{\rm eff}(\sigma)=V_0(\sigma)+\frac{1}{2\Omega_{\rm vol}}
             \ln\det\left\{\mu^{-2}\left(\Box_{(n)}
             -M^2_{\phi}\right)\right\}\,,
  \label{eq;3-2-8}
\end{equation}
where $\Box_{(n)}$ is the $n$-dimensional 
Laplace-Beltrami operator and $V_0(\sigma)$ is given by 
%
\begin{equation}
V_0(\sigma)=\frac{\bar{\Lambda}}{\kappa^2}
       \,{\rm e}^{\frac{-2d}{n-2}\frac{\sigma}{\sigma_{0}}}
 	-\frac{Kd(d-1)}{2\kappa^2b_{0}^2}
       \,{\rm e}^{\left(\frac{-2d}{n-2}+2\right)\frac{\sigma}{\sigma_{0}}}\,,
  \label{eq;3-2-9}
\end{equation}
with $\Omega_{\rm vol}$ the volume of the $n$-dimensional 
de Sitter spacetime. 
  
\subsection{Zeta function regularization}
       \label{subsec;zeta}
Now we need to evaluate the functional determinant appearing in the
 expression for the effective potential on a background manifold 
in $dS_n\times S^d$.
We apply the standard technique of zeta function regularization in order
 to deal with the ultraviolet divergence with respect to the 
 Kaluza-Klein mode.
 Many authors have previously used this method to study the
Casimir effect for the spacetime $M^4\times S^d$. 
\cite{can}${}^{,}$\cite{ki}${}^{-}$\cite{bi1}
We now define the functional determinants in terms of the generalized 
zeta function, which is the sum of operator eigenvalues, as  
%
\begin{equation}
\zeta_{n\times d}(s)\equiv\sum^{\infty}_{l=0}\sum^{\infty}_{l'=0}
                   D(l)\:d(l')\left\{\frac{\lambda(l')}{a^2}
                   +\left(\frac{b_0}{b}\right)^{2d/(n-2)}
                    \frac{\Lambda(l)}{b_0^2}\right\}^{-s}\,, 
  \label{eq;3-3-1}
\end{equation}
where $a$ and $b$ are the scales of $dS_n$ and $S^d$,  $\lambda(l')$ and 
$\Lambda(l)$ are eigenvalues of the scalar field $\phi$ in $dS_n$ and 
 $S^d$, and 
$d(l')$ and $D(l)$ are degeneracies of $dS_n$ and $S^d$, respectively.
This expansion is well-defined and converges for $Re(z)>(n+d)/2$.
Using this function, the effective potential (\ref{eq;3-2-8}) is written
 
%
\begin{equation}
V_{\rm eff}(\sigma)=V_0(\sigma)-\frac{1}{2\Omega_{\rm vol}}
             \left\{{\zeta'}_{n\times d}(0)
             +\zeta_{n\times d}(0)\ln({\mu}^2a^2)\right\}\,,      
  \label{eq;3-3-2}
\end{equation}
where, in order to get the second term, we have used the relation
%
\begin{equation}
\det\left(\mu M\right)=\mu^{\zeta(0)}\det M\,.
      \label{eq;zr}
\end{equation}
From this point, we drop the ``$n\times d$'' subscript on $\zeta$.
Our task is to calculate $\zeta(s)$ and to continue it analytically to
 $s=0$. ($\zeta$ is regular at $s=0$.) 
Note that the logarithmic term proportional to $\ln (\mu^2a^2)$ is  
ill-defined if a rescaling of $\mu$
 adds a linear combination of $\zeta(0)$ to it. 
It has been pointed out by several authors \cite{bi1}${}^{,}$ \cite{gi} 
that the finite term in the quantum effect depends
 on the regularity technique used. Although those authors were not able
 to  calculate $\zeta(0)$ for a technical reason, we evaluate the
 terms $\zeta'(0)$ and $\zeta(0)$ 
explicitly. (see Appendices \ref{sec:scalar},
 \ref{sec:spinor} and \ref{sec:vector}.)  

\subsection{The effective potential for $dS_4\times S^7$}
      \label{subsec;EP}  
In this subsection, we compute the effective potential of the dilaton
field coming from quantum effects of the background scalar, spinor,
vector and tensor fields
 in a background geometry that is the direct product of a
four-dimensional de Sitter spacetime with a seven-dimensional sphere.

 \vspace{0.3cm}
\subsubsection{Scalar field}

 \vspace{0.3cm}
The zeta function $\zeta$ can be evaluated if there is a formula for
 the eigenvalue of the operator in de Sitter spacetime
that is a four-dimensional hyperboloid of constant
 curvature.  It has a unique Euclidean section which is a four-sphere
 $S^4$ of radius $a$. The curvature tensors are 
%
\begin{equation}
R_{\mu\nu\rho\sigma}=\frac{k}{a^2}
        \left(g_{\mu\rho}g_{\nu\sigma}-g_{\mu\sigma}g_{\nu\rho}\right)\,,
    \hspace{1cm} 
R_{\mu\nu}=\frac{3k}{a^2}g_{\mu\nu}\,,
    \hspace{1cm}
R=\frac{12k}{a^2}\,,
\end{equation}
where $k$ is the curvature parameter for $dS_4$ and is  
taken as $k=1$. 
Fortunately, 
the degeneracy $d_{4}(l)$ and the eigenvalue $\lambda_{4}(l)$ of the 
massless scalar field in $dS_4$ spacetime 
are known. They are given by 
%
\begin{equation}
d_{4}(l)=\frac{1}{6}(l+1)(l+4)(2l+3)\,,\hspace{1cm}
\lambda_{4}(l)=\frac{1}{a^2}l(l+3),\hspace{1cm} l=0,\:1,\:2,\cdots\,,
\end{equation}
where $a$ is the radius of the four-sphere.
It is necessary to regularize 
$\zeta(s)$ in order to evaluate the effective potential $V_{\rm eff}$
 because $\zeta(s)$ has a divergence resulting from 
the infinite mode sum. 
We use the prescription of the zeta function regularization 
given by Kikkawa et al.\cite{ki}  In Appendix \ref{sec:scalar}, 
we give the details of the calculation of the zeta function in the massive
 scalar field case.
To simplify the calculation, it is assumed that the 
scale of de Sitter spacetime, $a$, is much larger than
that of the compactification scale, $b$.
Under this assumption, $V_{\rm eff}$ is
 evaluated by numerical integration of  
 several terms appearing in $\zeta'(0)$ and $\zeta(0)$.      
We can see that the terms proportional to 
$(a/b_0)^4(b_0/b)^{18}$ in Eq.\,(\ref{eq;a-19})
 are dominant when the condition $a \gg b_0$ is satisfied.
 Our final result for the scalar
 potential to 1-loop order is then 
%
\begin{eqnarray}
V_{\rm eff}(\sigma)&=&V_0(\sigma)
        -\frac{1}{2\Omega_{\rm vol}}\left\{\zeta'(0)
        +\zeta(0)\ln(\mu^2a^2)\right\}\nonumber\\
       &=&\frac{\Lambda}{\kappa^2}\,{\rm e}^{-7\frac{\sigma}{\sigma_0}}
       -\frac{21K}{\kappa^2 b_0^2}\,{\rm e}^{-9\frac{\sigma}{\sigma_0}}
       +\frac{2.61083\times 10^{-2}}{b_0^4}\,
        {\rm e}^{-18\frac{\sigma}{\sigma_0}}\,,
     \label{eq;potential}
\end{eqnarray}
where $V_0(\sigma)$ is given by Eq.\,(\ref{eq;3-2-9}).

 \vspace{0.3cm}
\subsubsection{Dirac Spinor field}

 \vspace{0.3cm}
Next, we calculate 
the quantum effect associated with the massless Dirac spinor field $\psi$
 on $dS_4\times S^7$
with the action 
%
\begin{equation}
I_{\psi}=i\int d^{11}x\;\:\sqrt{-\bar{g}}\bar{\psi} 
                   \:\bar{\gamma}^M\:\bar{\nabla}_M\psi\,.
\end{equation}
The eleven-dimensional gamma matrix $\bar{\gamma}$ is given by\cite{can} 
%
\begin{equation}
\bar{\gamma}^{\mu}=\gamma^{\mu}\otimes {\bf 1}\,,\hspace{1cm}
\bar{\gamma}^i=\gamma^5\otimes\gamma^i\,,\hspace{1cm}
\left(\gamma^5\right)^2=1\,,\hspace{1cm}
\left\{\bar{\gamma}^M\,,\;\bar{\gamma}^N\right\}=2g^{MN}\,,
\end{equation}
where the $\gamma^{\mu}\:(\mu=0,\:1,\:2,\:3)$ are the Dirac matrices in 
$dS_4$, while the $\gamma^i\:(i=4,\:5,\:\cdots,\:10)$ are those in $S^7$.
The spinor representation of $O(1, 3+7)$ is a direct product of the
spinor representation of $O(1, 3)$ and $O(7)$.
The Dirac spinor field $\psi$ is expanded in spinor harmonics,
 in analogy to the scalar field:\cite{gil}
%
\begin{equation}
 \bar{\psi}(x^{\mu},\;x^i) = b_{0}^{-7/2}
	\sum_{l,m}\psi_{lm}(x^{\mu})Y^{(7)}_{\psi\;lm}(x^i)\,.
   \label{eq;3-fh}
\end{equation}
Here, the constant $b_{0}$ is the initial value of $b$, and 
$\psi_{lm}(x^{\mu})$ is the Dirac spinor field in the
 four-dimensional spacetime.
In Eq.\,(\ref{eq;3-fh}), $Y^{(7)}_{\psi\;lm}$ 
(where $l=1,2,\cdots$, and $m$ denotes a set of six  
numbers that are needed in order for the set of all 
$Y^{(7)}_{\psi\;lm}$ to be a complete set of $L^2$ functions on 
 $S^7$) are real spinor harmonics on $S^7$ satisfying 
%
\begin{eqnarray}
i\:\bar{\gamma}^i\nabla_i\;Y^{(7)}_{\psi\;lm}
 & = &\Lambda_{\psi}Y^{(7)}_{\psi\;lm}\,,\\
 \int d^7x\sqrt{\Omega^{(7)}}\:Y^{(7)}_{\psi\;lm}\:Y^{(7)}_{\psi\;l'm'} 
 & = & \delta_{ll'}\delta_{mm'}\,,
\end{eqnarray}
and $\psi_{lm}(x^{\mu})$ is a real function depending only on 
the four-dimensional coordinates $x^{\mu}$. 
Here, $\bar{\gamma}^i\nabla_i$ is the Dirac operator on the unit seven 
sphere 
$S^7$, and $\Lambda_{\psi}(l)$ denotes an eigenvalue for the Dirac spinor
field $\psi$.
Using the relation for the eleven-dimensional Dirac operator 
$\bar{\gamma}^M\nabla_M=\bar{\gamma}^{\mu}\nabla_{\mu}\otimes {\bf 1}
                       +\gamma^5\otimes\bar{\gamma}^i\nabla_i$ and the 
conformal transformation (\ref{eq;2-4}), 
 we obtain the four-dimensional effective action 
%
\begin{equation} 
I\left(\psi_{lm},\;\bar{\psi}_{lm}\right)  \sum_{l,m}\int d^4x \:\sqrt{-g}
   \,{\rm e}^{-7\sigma/\sigma_0}\bar{\psi}_{lm}
  \left(i\gamma^{\mu}\nabla_{\mu}+\Lambda_{\psi}\:\gamma^5\right)
  \psi_{lm}\,.
\end{equation}
The partition function $Z$ for the 
massless Dirac spinor field on $dS_4$ is 
%
\begin{equation}
Z=\int {\cal D}\psi_{lm}\,{\cal D}\bar{\psi}_{lm}
 \exp\left\{-iS\left(\psi_{lm},\;\bar{\psi}_{lm}\right)\right\}\,,
   \label{eq;fermi-par}
\end{equation}
where ${\cal D}\psi_{lm}$ and 
 ${\cal D}\bar{\psi}_{lm}$ are the functional measures over
the spinor field $\psi_{lm}$ and its Dirac adjoint field 
$\bar{\psi}_{lm}$, respectively.
Using the definition of a Gaussian functional for anti-commuting fields 
 Eq.\,(\ref{eq;fermi-par}) become
%
\begin{equation}
Z=\det\left\{\mu^{-1}\,{\rm e}^{-7\frac{\sigma}{\sigma_0}}
  \left(i\gamma^{\mu}\nabla_{\mu}
  +\Lambda_{\psi}(l)\gamma^5\right)\right\}\,.
\end{equation}
Then, we obtain 
%
\begin{eqnarray}
\ln Z&=&\ln\det\left\{\mu^{-1}
  \left(i\gamma^{\mu}\nabla_{\mu}
  +{\rm e}^{-7\frac{\sigma}{\sigma_0}}
   \Lambda_{\psi}(l)\gamma^5\right)\right\}\nonumber\\
 &=&\frac{1}{2}\ln\det\left[\mu^{-2}
   \left\{-\left(\gamma^{\mu}\nabla_{\mu}\right)^2
  +{\rm e}^{-14\frac{\sigma}{\sigma_0}}
  \left(\Lambda_{\psi}(l)\right)^2\right\}\right]\,,
\end{eqnarray}
where the four-dimensional Dirac operator 
$\left(\gamma^{\mu}\nabla_{\mu}\right)^2$ is given by 
\cite{sc}${}^{,}$\cite{li}
%
\begin{equation}
\left(\gamma^{\mu}\nabla_{\mu}\right)^2=\Box+\frac{1}{4}R_{(4)}\,,
\end{equation}
and $\Box$ denotes the four-dimensional Laplace-Bertlami operator. 
Following the procedure of $\S\S$\ref{subsec;quantum} and
\ref{subsec;zeta}, the effective potential must
satisfy the relation 
%
\begin{equation}
V_{\rm eff}(\sigma)=V_0(\sigma)
         -\frac{1}{4\Omega_{\rm vol}}\left\{\zeta'(0)
           +\zeta(0)\ln(\mu^2a^2)\right\}\,,
\end{equation}
where $\zeta(0)$, which is defined in Appendix \ref{sec:spinor},
 is the generalized zeta function for the Dirac 
spinor field and $V_0(\sigma)$ is expressed by Eq.\,(\ref{eq;3-2-9}). 
We can see that under the assumption $a\gg b_0$,  
the dominant term in $\zeta'(0)$ is proportional to 
${\rm e}^{-18\sigma/\sigma_0}$. 
Using the calculation method presented in Appendix \ref{sec:spinor}, 
the effective potential to 1-loop order is finally given by 
%
\begin{equation}
V_{\rm eff}(\sigma)=\frac{\Lambda}{\kappa^2}
       \,{\rm e}^{-7\frac{\sigma}{\sigma_0}}
       -\frac{21K}{\kappa^2 b_0^2}\,{\rm e}^{-9\frac{\sigma}{\sigma_0}}
       -\frac{1.258212\times 10^{-6}}{b_0^4}\,
        {\rm e}^{-18\frac{\sigma}{\sigma_0}}\,.
        \label{eq;fpotential}
\end{equation}

 \vspace{0.3cm}
\subsubsection{$U(1)$ gauge field}

 \vspace{0.3cm}
We now compute the quantum correction for the vector field $A_M$ in
eleven dimensions.
We take the eleven-dimensional action for $U(1)$ gauge field 
to be
%
\begin{equation}
I_{U(1)}=\frac{1}{4}\int d^{11}x\:\sqrt{-\bar{g}}\;F_{MN}F^{MN}\,,
     \label{eq;3c-1}
\end{equation}
where $F_{MN}=\nabla_{M}A_N-\nabla_{N}A_M$.
In order to perform the dimensional reduction for 
 the $U(1)$ field action in $dS_4\times S^d$
 spacetime, it is convenient to
 expand in the vector harmonics on $S^7$ as 
%
\begin{equation}
\bar{A}_Mdx^M=b_0^{-\frac{7}{2}}\sum_{l,\;m}
      \left[A^{(4)}_{\mu\;lm}Y^{(7)}_{lm}
      dx^{\mu}+\left\{A^{(4)}_{{\rm (T)}\;lm}
     \left(V^{(7)}_{{\rm (T)}\;lm}\right)_i
  +A^{(4)}_{{\rm (L)}\;lm}
\left(V^{(7)}_{{\rm (L)}\;lm}\right)_i\right\}dx^i\right]\,,
     \label{eq;3c-2}
\end{equation}
where $A^{(4)}_{\mu\;lm}$, $A^{(4)}_{{\rm (T)}\;lm}$
 and $A^{(4)}_{{\rm (L)}\;lm}$
depend only on the four-dimensional
coordinates ${x^{\mu}}$, and $Y^{(7)}_{lm}$, 
$V^{(7)}_{{\rm (T)}\;lm}$ and  
$V^{(7)}_{{\rm (L)}\;lm}$ are the scalar
harmonics, transverse vector harmonics and longitudinal vector
harmonics, respectively. 
[see the Appendix B of Ref.\,36)] for definitions 
and properties of there harmonics.)
The summations are taken over values of $l$ satisfying 
$l\ge 0$ for the scalar harmonics
and $l\ge 1$ for the vector harmonics. 
This decomposition is unique and orthogonal.    
As $A^{(4)}_{{\rm (L)}\;lm}$ represents the gauge degrees of freedom,
  we eliminate
 the longitudinal mode for $A_M$ after the gauge fixing. [see also the
 appendix in the Ref.\,36).] 
By substituting the expansion (\ref{eq;3c-2}) into the action
(\ref{eq;3c-1}) and using the conformal
transformation (\ref{eq;2-4}), we obtain the four-dimensional 
effective action\cite{uz1}
%
\begin{eqnarray}
I_{U(1)} & = & I_{\rm (T)}+I_{\rm (V)}\,,
          \label{eq;3c-t}
        \nonumber\\
 I_{\rm (T)} & = & -\frac{1}{2}\int dx^4\sqrt{-g}
	\left[{\rm e}^{-2\frac{\sigma}{\sigma_0}}g^{\mu\nu}
	\partial_{\mu}A_{\rm (T)}\partial_{\nu}A_{\rm (T)}
	+{\rm e}^{-11\frac{\sigma}{\sigma_0}}
	M^2_{\rm (T)}A_{\rm (T)}^2\right]\,,
	\nonumber\\
 I_{\rm (V)} & = & -\int dx^4\sqrt{-g}\left[
	\frac{1}{4}{\rm e}^{7\frac{\sigma}{\sigma_0}}
	g^{\mu\rho}g^{\nu\sigma}F_{\mu\nu}F_{\rho\sigma}
	+ \frac{1}{2}{\rm e}^{-2\frac{\sigma}{\sigma_0}}
	M_{\rm (V)}^2g^{\mu\nu}A_{\mu}A_{\nu}\right]\,,
	\label{eq;3c-3}
\end{eqnarray}
where $F_{\mu\nu}=\partial_{\mu}A_{\nu}-\partial_{\nu}A_{\mu}$. 
The terms containing $M_{\rm (T)}^2$ and $M_{\rm (V)}^2$ in the 
$U(1)$ gauge field action,
given by  
%
\begin{eqnarray}
M^2_{\rm (T)}=\frac{l(l+6)+5}{b_0^2}\,, \hspace{1cm}
M_{\rm (V)}^2=\frac{l(l+6)}{b_0^2}\,.
     \label{eq;3c-4}
\end{eqnarray}
denote the mass of the four-dimensional scalar field $A_{\rm (T)}$ 
and the vector field $A_{\mu},$ respectively.
The vector field $A_{\mu}$ can be decomposed with 
Hodge decomposition, which is the same as $A_M$.
We expand the field 
$A_{\mu}$ in terms of harmonics in the four-dimensional de Sitter
 spacetime of constant curvature as 
%
\begin{equation}
A^{(4)}_{\mu}dx^{\mu}=\left(A^{(4)}_{{\rm (T)}\:\mu}
              +A^{(4)}_{{\rm (L)}\;\mu}\right)dx^{\mu}\,.
     \label{eq;3c-5}
\end{equation}
Here, we consider a gauge fixing for the four-dimensional $U(1)$ field 
$A^{(4)}_{\mu}$.
A simple and natural gauge choice for the quantization of this system 
is the Lorentz gauge, $\nabla^{\mu}A_{\mu}=0$.
The total action (\ref{eq;3c-t})
 can be affected by the addition of the gauge fixing action
%
\begin{equation}
\delta I=\frac{1}{2}\alpha\int d^4x\;\sqrt{-g}
 \left(\nabla_{\mu}A^{\mu}\right)^2\,, 
     \label{eq;3c-6}
\end{equation}
where $\alpha$ is a positive constant. 
The De Witt-Fadeev-Popov ghost factor $\Delta$ is defined by 
%
\begin{equation}
\Delta\int {\cal D}\eta\exp(-\delta I)=1\,,
     \label{eq;3c-7}
\end{equation}
where ${\cal D}\eta$ is the measure on the gauge group. 
The ghost factor must be independent of the gauge field. 
We can define the elements of the gauge group as
$\eta=\exp(i\omega(x))$.\cite{al} 
The field $\omega(x)$ is written in terms of
 orthogonal scalar eigenfunctions
 as $\omega(x)=\sum^{\infty}_{n=0}\omega_n\chi_n(x)$.   
These elements transform the gauge fields
$A_{\mu }$ into $A_{\mu}+\partial_{\mu}\omega(x)$.  
The action is invariant under this $U(1)$ gauge transformation. 
The measure of the $U(1)$ gauge group ${\cal D}\eta$ is 
then written 
%
\begin{equation}
{\cal D}\eta=\prod^{\infty}_{n=1}\mu^2d\omega_n\,.
     \label{eq;3c-a}
\end{equation}
Using these gauge elements, the ghost factor $\Delta$ is explicitly 
expressed as 
%
\begin{equation}
\Delta=\mu^{-2}a^{-2}\det\left(\alpha^{1/2}\mu^{-2}\bar{Q}\right)\,,
     \label{eq;3c-8}
\end{equation}
where $\bar{Q}$ is an operator for scalar fields without a zero mode. 
The total amplitude in the path integral is expressed by 
%
\begin{equation}
Z=\Delta\int {\cal D}A^{(4)}_{\rm (T)}
             {\cal D}A^{(4)}_{{\rm (T)}\;\mu}
  {\cal D}\eta
  \exp\left\{-\left(I\left[A^{(4)}_{\rm (T)},
  \;A^{(4)}_{{\rm (T)}\;\mu}\right]
  +\delta I\right)\right\}\,.
     \label{eq;3c-9}
\end{equation}
Using the action given in Eqs.\,(\ref{eq;3c-3}) and (\ref{eq;3c-6}), 
we compute the quantum correction for the $U(1)$ gauge field.
 However, this action is not useful for the calculation 
of quantum correction to vector field due to the coupling dilaton.
After integrating the kinetic terms of $A^{(4)\;\mu}_{{\rm (T)}\;\mu}$
 and $A^{(4)}_{{\rm (T)}\;\mu}$ by parts,  the actions are given by 
%
\begin{eqnarray}
\hspace{-1cm} 
 I_{\rm (V)} & = & -\sum_{l,\;m}\frac{1}{2}\int dx^4\;\sqrt{-g}\;
	A^{(4)\;\mu}_{{\rm (T)}\;lm}\left\{\frac{1}{2}
        \left(g_{\mu\nu}\Box-\nabla_{\mu}\nabla_{\nu}\right)
        {\rm e}^{7\frac{\sigma}{\sigma_0}}\right. \nonumber\\
       & &\left. 
        -\,{\rm e}^{7\frac{\sigma}{\sigma_0}}\bigtriangleup_{\mu\nu}
        +g_{\mu\nu}\,{\rm e}^{-2\frac{\sigma}{\sigma_0}}
         M^2_{\rm (V)}\right\}A^{(4)\;\nu}_{{\rm (T)}\;lm}\,,
	\\
\hspace{-1cm}
 I_{\rm (T)} & = & -\sum_{l,\;m}\frac{1}{2}\int dx^4\sqrt{-g}
        A^{(4)}_{{\rm (T)}\;lm}\left\{
        \frac{1}{2}\Box\,{\rm e}^{-2\frac{\sigma}{\sigma_0}}
        -{\rm e}^{-2\frac{\sigma}{\sigma_0}}\Box+M^2_{\rm (T)}
        \right\}A^{(4)}_{{\rm (T)}\;lm}\,,
	\label{eq:3c-11}
\end{eqnarray}
where $\bigtriangleup_{\mu\nu}=g_{\mu\nu}\Box+R_{\mu\nu}$.
The total action can be expressed by  
%
\begin{eqnarray}
I+\delta I&=&\frac{1}{2}\int d^4x 
\left[A^{(4)\;\mu}_{{\rm (L)}\;lm}
      \left(\alpha\nabla_{\mu}\nabla_{\nu}\right)
       A^{(4)\;\nu}_{{\rm (L)}\;lm}
       \right.\nonumber\\
     &+&A^{(4)\;\mu}_{{\rm (T)}\;lm}\left\{\frac{1}{2}
        \left(g_{\mu\nu}\Box-\nabla_{\mu}\nabla_{\nu}\right)
        \,{\rm e}^{-7\frac{\sigma}{\sigma_0}}
        -{\rm e}^{-7\frac{\sigma}{\sigma_0}}\bigtriangleup_{\mu\nu}
        +g_{\mu\nu}M^2_{\rm (V)}
      \right\}A^{(4)\;\nu}_{{\rm (T)}\;lm}
       \nonumber\\
     &+&\left.A^{(4)}_{{\rm (T)}\;lm}\left\{
     \frac{1}{2}\Box\,{\rm e}^{-2\frac{\sigma}{\sigma_0}}
        -{\rm e}^{-2\frac{\sigma}{\sigma_0}}\Box+M^2_{\rm (T)}
      \right\}
      A^{(4)}_{{\rm (T)}\;lm}\right]\,,
     \label{eq;3c-13}
\end{eqnarray}
where we ignore terms of higher than more than quadratic order 
in the field.
From the above action, we find the functional $Z$ to be
%
\begin{eqnarray}
Z&=&\mu^{-2}a^{-2}\det\left(\alpha^{1/2}\mu^2Q_{\rm (L)}\right)
\det\left(\mu^2Q_{\rm (T)}\right)^{1/2}
\det\left(\mu^2Q_{\rm (S)}\right)^{1/2} 
\det\left(\alpha\mu^2Q_{\rm (L)}\right)^{-1/2}\nonumber\\
 &=&\mu^{-2}a^{-2}\det\left(\mu^2Q_{\rm (L)}\right)^{1/2}
\det\left(\mu^2Q_{\rm (T)}\right)^{1/2}
\det\left(\mu^2Q_{\rm (S)}\right)^{1/2}\,, 
\end{eqnarray}
where  we have absorbed a numerical constant into $\mu$,
 and $Q_{\rm (S)}$ denotes the operator for scalar field
 $A_{\rm (T)}$, and 
$Q_{\rm (T)}$ and $Q_{\rm (L)}$ denote the operators for 
vector fields $A_{{\rm (T)}\;\mu}$ and $A_{{\rm (L)}\;\mu}$, 
respectively. Note that
$\alpha$ has cancelled out in $z$.
Using the zeta function regularization, 
we can compute the effective
potential for the vector field in the $dS_4\times S^7$ background. 
We do not consider the contribution from
 $A^{(4)}_{{\rm (L)}\;lm}$, because 
it is not dominant term in the effective potential. 

 \vspace{0.3cm}
%
\begin{center} 
{\bf (i) 1-loop quantum correction from ${\rm A^{(4)}_{(T)}}$}
\end{center}

 \vspace{0.3cm}
First, we consider the field $A^{(4)}_{\rm (T)}$. This is 
a scalar field on $dS_4$ [see Eq.\,(\ref{eq;3c-2})].
We define the generalized zeta function
%
\begin{equation}
\zeta_{\rm T}(z)\equiv\sum^{\infty}_{l=0}\sum^{\infty}_{l'=0}
                   D_{\rm (T)}(l)\:d_{\rm (T)}(l')
                   \left\{{\rm e}^{-2\frac{\sigma}{\sigma_0}}
                   \lambda^2_{\rm (T)}(l')a^{-2}
         +\Lambda^2_{\rm (T)}(l)\,{\rm e}^{-11\frac{\sigma}{\sigma_0}}
                   \right\}^{-z}\,, 
     \label{eq;3c-15}
\end{equation}
where $d_{\rm (T)}(l')$ and $D_{\rm (T)}(l)$, given by  
%
\begin{equation}
D_{\rm (T)}(l)=\frac{(2l+5)(l+4)!}{720l}\,,\hspace{1.5cm}
d_{\rm (T)}(l')=\frac{(2l+3)(l+2)!}{6l}\,, 
     \label{eq;3c-17}
\end{equation}
denote the degeneracies of 
$A^{(4)}_{\rm (T)}$ in $dS_4$ and $ S^7$, respectively, and 
$\Lambda^2_{\rm (T)}(l)$ and $\lambda^2_{\rm (T)}(l')$, given by
%
\begin{equation}
\Lambda^2_{\rm (T)}(l)=\frac{l(l+6)+5}{b_0^2}\,,\hspace{1.5cm}
\lambda^2_{\rm (T)}(l')=\frac{1}{2}l(l+3)\,, 
     \label{eq;3c-18}
\end{equation}
denote the eigenvalues for $A^{(4)}_{\rm (T)}$ in $dS_4$ and $S^7$,
 respectively.
Using the same strategy as in the case of the scalar and spinor fields,
 we can regularize the $\zeta$ function for the vector
field. 
Then, we can compute $\zeta_{\rm T}'(0)$ and $\zeta_{\rm T}(0)$ and
 pick up the dominant term in $\zeta_{\rm T}'(0)$ and
 $\zeta_{\rm T}(0)$.     
In this way, we find that the dominant term in the effective potential 
for $A^{(4)}_{\rm (T)}$ is proportional to
$b^{-18}$ for $A^{(4)}_{\rm (T)}$.
%

 \vspace{0.3cm}
%
\begin{center}
{\bf (ii) 1-loop calculation for ${\rm A^{(4)}_{(T)\;\mu}}$}
\end{center}

 \vspace{0.3cm}
Next, we compute the quantum effect from $A^{(4)}_{{\rm (T)}\;\mu}$.
 This consists of the vector field components in $dS_4$
 [see Eq.\,(\ref{eq;3c-2})]. 
We define the zeta function as
%
\begin{equation}
\zeta_{\rm V}(z)\equiv\sum^{\infty}_{l=0}\sum^{\infty}_{l'=0}
                  D_{\rm (V)}\:(l)\:d_{\rm (V)}(l')\left\{
                  {\rm e}^{-7\frac{\sigma}{\sigma_0}}
                 \lambda^2_{\rm (V)}(l')a^{-2}
                 +\Lambda^2_{\rm (V)}(l)\;b^{-2}\right\}^{-z}\,, 
\end{equation}
where $D_{\rm (V)}\:(l)$ and  $d_{\rm (V)}(l)$, given by
%
\begin{equation}
D_{\rm (V)}(l)=\frac{(2l+5)(l+4)!}{5!l}\,,\hspace{1.5cm}
d_{\rm (V)}(l')=\frac{(2l+3)(l+2)!}{3!l}\,, 
\end{equation}
denote the degeneracies of the vector fields 
in $S^7$ and $dS_4$, and 
$\Lambda_{\rm (V)}^2(l)$ and $\lambda_{\rm (V)}^2(l')$, given by  
%
\begin{equation}
\Lambda^2_{\rm (V)}(l)=\frac{1}{2}l(l+6)\,,\hspace{1.5cm}
\lambda^2_{\rm (V)}(l')=l(l+3)-\frac{5}{2}\,,
\end{equation}
denote the eigenvalues for the vector fields 
in  $S^7$ and $dS_4$,\cite{ru1} respectively.
Using a formula in Appendix \ref{sec:vector}, 
we see that $\zeta_{\rm V}(0)$ is also vanishing, and the dominate term 
in $\zeta_{\rm V}'(0)$ is proportional to $b^{-18}$. 
Finally, we obtain the quantum correction for the 
$U(1)$ vector field as
%
\begin{eqnarray}
V_{\rm eff}({\sigma})&=&V_0(\sigma)-\frac{1}{2\Omega_{\rm vol}}\left[
        \left\{\zeta_{\rm T}'(0)+\zeta_{\rm V}'(0)\right\}        
        +\left\{\zeta_{\rm T}(0)+\zeta_{\rm V}(0)\right\}
        \ln(\mu^2a^2)\right]\nonumber\\
       &=&\frac{\Lambda}{\kappa^2}\,{\rm e}^{-7\frac{\sigma}{\sigma_0}}
       -\frac{21K}{\kappa^2 b_0^2}\,{\rm e}^{-9\frac{\sigma}{\sigma_0}}
       +\frac{2.1974\times 10^{-3}}
       {b_0^4}\,{\rm e}^{-18\frac{\sigma}{\sigma_0}}\,,
        \label{eq;vpotential}
\end{eqnarray}
where $V_0(\sigma)$ is given by Eq.\,(\ref{eq;3-2-9}).
 
 \vspace{0.3cm}
\subsubsection{Tensor field}

 \vspace{0.3cm}
Next, we calculate the quantum correction of the gravitational
 field. Although a full 
quantum theory of gravity does not exist, the quantum gravitational effect
at the
 1-loop level is important for the Kaluza-Klein theories, because the
scale of the
 internal spacetime in the Kaluza-Klein picture is assumed to be not far
larger than the Planck length.

 We consider a gravitational perturbation $h_{MN}$ around a
background metric $\bar{g}_{MN}^{(0)}$, which we specify below: 
%
\begin{equation}
\bar{g}_{MN} =\bar{g}_{MN}^{(0)}+h_{MN}\,.
\label{eqn;2}
\end{equation}
First,    
by substituting Eq.\,(\ref{eqn;2}) into Eq.\,(\ref{eq;2-1}),
 we obtain the
perturbed Einstein-Hilbert action as
%
\begin{eqnarray}
 I_{\rm EH} & =  & \frac{1}{2\bar{\kappa}^2} 
 	\int d^{11}x\sqrt{-\bar{g}^{(0)}} 
	\left[ \bar{R}^{(0)} -2\bar{\Lambda}
        -h^{MN}\left(\bar{R}^{(0)}_{MN}
	-\frac{1}{2}\bar{R}^{(0)}\bar{g}^{(0)}_{MN}
	+\bar{\Lambda} \bar{g}^{(0)}_{MN}\right)\right.\nonumber\\
 & & + \frac{1}{8}\left( h^2-2h^{MN}h_{MN}\right)\bar{R}^{(0)} 
	+\frac{1}{2}\left(2h^{MM'}h_{M'}^N-hh^{MN}\right)
	\bar{R}^{(0)}_{MN}  \nonumber  \\ 
 & &  +\frac{1}{4} \left\{ h^{MN}_{\quad;M'}
	\left(2h^{M'}_{M;N}- h_{MN}^{\quad;M'}\right) 
	+ h_{;M} \left(h^{;M}-2h^{MN}_{\quad;N}\right) 
	\right\}\nonumber\\
 & & \left.
	-\bar{\Lambda}\left(\frac{1}{4}h^2-\frac{1}{2}h^{MN}h_{MN} 
	\right)  +O(h^3)\right]\,,
	\label{eqn:perturbedEH}
\end{eqnarray}
where ``$;$'' denotes the covariant derivative compatible with
$g^{(0)}_{MN}$, and $\bar{R}^{(0)}_{MN}$ and $\bar{R}^{(0)}$ are 
the Ricci tensor and scalar constructed from
$\bar{g}^{(0)}_{MN}$.
For the background geometry $dS_4\times S^7$,
 we compactify this action on the seven-dimensional sphere $S^7$. 
The gravitational perturbation $h_{MN}$ in this particular background
 can be expanded in harmonics on $S^7$ as 
%
\begin{eqnarray}
 h_{MN}dx^Mdx^N &=& \sum_{l,\:m}
 	\left[ h_{\mu\nu}^{lm}
	Y_{lm}dx^{\mu}dx^{\nu}  
 	+2 \{h_{{\rm (T)}\mu}^{lm}(V_{{\rm (T)}lm})_i
	+ h_{{\rm (L)}\mu}^{lm}(V_{{\rm (L)}lm})_i \}
	dx^{\mu}dx^i \right.\nonumber \\
 & & +
	\{ h_{\rm (T)}^{lm}(T_{{\rm (T)}lm})_{ij}
	+ h_{\rm (LT)}^{lm}(T_{{\rm (LT)}lm})_{ij}
	+ h_{\rm (LL)}^{lm}(T_{{\rm (LL)}lm})_{ij}\nonumber \\
 & & + \left.
	+ h_{\rm (Y)}^{lm}(T_{{\rm (Y)}lm})_{ij}\}
	dx^idx^j \right]\,,\nonumber\\
       \label{eq;3D-3}
\end{eqnarray}
where the $Y_{lm}$ are 
the scalar harmonic functions, $V_{{\rm (T)}lm}$ and
$V_{{\rm (L)}lm}$ are the vector harmonics, and $T_{{\rm (T)}lm}$,
 $T_{{\rm (LT)}lm}$,
 and $T_{{\rm (LL)}lm}$ are the tensor harmonics. Here, the 
coefficients $h_{\mu\nu}^{lm}$, $h_{{\rm (T)}\mu}^{lm}$,
 $h_{{\rm (L)}\mu}^{lm}$, 
$h_{\rm (T)}^{lm}$, $h_{\rm (LT)}^{lm}$, $h_{\rm (LL)}^{lm}$
 and $h_{\rm (Y)}^{lm}$
depend only on the four-dimensional coordinates $x^{\mu}$, while the
harmonics depend only on the coordinates $x^i$ on $S^7$. 

Although the above
 expression of $h_{MN}$ includes many terms, some of them 
represent degrees of freedom of coordinate transformations. In fact,
it is shown in appendix of Ref.\,36) that,
 after gauge-fixing and
redefining $g_{\mu\nu}$ and $b$, the perturbation $h_{MN}$ can be
expressed as 
%
\begin{eqnarray}
 h_{MN}dx^Mdx^N &=& \sum_{l,\:m}
	\left[ h_{\mu\nu}^{lm}Y_{lm}dx^{\mu}dx^{\nu}
	+ 2h_{{\rm (T)}\mu}^{lm}(V_{{\rm (T)}lm})_idx^{\mu}dx^i
        \right.\nonumber \\
        & &\left.+ \left\{ h_{\rm (T)}^{lm}(T_{{\rm (T)}lm})_{ij}
	+ h_{\rm (Y)}^{lm}(T_{{\rm (Y)}lm})_{ij}\right\}dx^idx^j\right]\,,
	\label{eqn:gaugefixedh}
\end{eqnarray}
where the summations are taken over values $l$ satisfying 
$l\ge 1$ for the scalar and vector 
harmonics and $l\ge 2$ for the tensor harmonics. 

Finally, by substituting this expression into
Eq.~(\ref{eqn:perturbedEH}), we obtain the action 
%
\begin{equation}
 I_{\rm EH} = I^{(0)}+I^{(1)}+I^{(2)}+O(h^3)\,,
	\label{eqn:perturbedEH2}
\end{equation}
where 
%
\begin{eqnarray}
& &I^{(0)}=\int d^4x\sqrt{-{g}^{ (0)}}\left[\frac{1}{2{\kappa}^2}R^{ (0)}
-\frac{1}{2}g^{\mu\nu}\partial_{\mu}\sigma\partial_{\nu}\sigma
-U_0(\sigma)\right], \\
& &I^{(2)}=\sum_{l, m}\int d^4x\sqrt{-{g}^{ (0) }}
\left({\cal L}^{\rm (T)}_{lm}+ {\cal L}^{\rm (V)}_{lm}
+ {\cal L}^{\rm (Y)}_{lm} \right)\,,
             \label{eq;3D-6}
\end{eqnarray}
and $I^{(1)}$ is linear in $h$. 
When the total action of the system is considered, $I^{(1)}$ should be
 canceled 
by other linear terms in the total action, because of the equations of
motion. 
Finally, the Lagrangian densities ${\cal L}^{\rm (T,V,Y)}_{lm}$ are given
by 
%
\begin{eqnarray}
{\cal L}^{\rm (T)}_{lm}&=&-\frac{1}{2}e^{-4\frac{\sigma}{\sigma_0}}
g^{\mu\nu}\partial_{\mu}\chi^{lm}\partial_{\nu}\chi^{lm}
-\frac{1}{2}M^{2}_{(\chi)\;lm}\chi^{lm}\chi^{lm}, \\
{\cal L}^{\rm (V)}_{lm}&=&{\cal L}^{\rm (V)}_{lm}
[h^{lm}_{{\rm (T)}\;\mu}],\\
{\cal L}^{\rm (Y)}_{lm}&=&{\cal L}^{\rm (Y)}_{lm}[\;h^{lm}_{\mu\nu},
h^{lm}_{\rm (Y)}\;]\,,
       \label{eq;3D-10}
\end{eqnarray}
where
%
\begin{eqnarray}
\hspace{-1cm}
 M^{2}_{(\chi)\;lm} & = &
	{\rm e}^{-13\frac{\sigma}{\sigma_0}}\{l(l+6)+30\}
	b_0^{-2}\nonumber\\
	&+&{\rm e}^{-4\frac{\sigma}{\sigma_0}}\left[11\nabla^2
	\left(\frac{\sigma}{\sigma_0}\right)
	-\frac{55}{2}\left\{\nabla\;
	\left(\frac{\sigma}{\sigma_0}\right)\right\}^2 
	+\left\{R^{(0)}-2e^{-7\frac{\sigma}{\sigma_0}}
         \bar{\Lambda}\right\}
	 \right]\,,\\
\hspace{-1cm}
 \chi^{lm} & \equiv & \sqrt{\frac{{b_0}^{-4}}{2^{d+2}\pi{\kappa}^2}}
	\;h^{lm}_{\rm (T)}\,. \qquad (l\ge 2) 
       \label{eq;3D-11}
\end{eqnarray}
[Since hereafter we analyze $h_{\rm (T)}^{lm}$ (or $\chi^{lm}$) only, we
have not included the explicit forms of ${\cal L}^{\rm (V)}_{lm}$ and
${\cal L}^{\rm (Y)}_{lm}$ here.] Hence, up to the second order with
 respect to the Kaluza-Klein
modes, $h_{\rm (T)}^{lm}$ (or $\chi^{lm}$) is decoupled from all other
Kaluza-Klein modes.   In this paper, for simplicity,
we investigate the quantum effects of $\chi^{lm}$ (or
$h_{\rm (T)}^{lm}$) only. 

Hereafter, we investigate the dynamics of the field $\chi^{lm}$
coupled to the dilaton field $\sigma$. These
fields are described by the action
%
\begin{equation}
I = I_{\sigma} + I_{\chi}\,,
\end{equation}
where
%
\begin{eqnarray}
I_{\sigma} & = & -\int d^4x\sqrt{-\bar{g}^{ (0)}}\left[
\frac{1}{2}g^{\mu\nu}\partial_{\mu}\sigma\partial_{\nu}\sigma
+U_0(\sigma)\right]\,,\nonumber\\
I_{\chi} & = & -\frac{1}{2}\int d^4x\sqrt{-\bar{g}^{ (0)}}\left[
{\rm e}^{-4\frac{\sigma}{\sigma_0}}
g^{\mu\nu}\partial_{\mu}\chi^{lm}\partial_{\nu}\chi^{hm}
+\Lambda^{\rm (T)}_{lm}\chi^{lm}\chi^{lm}\right]\,.
       \label{eq;3D-12}
\end{eqnarray}
Here, we mention that in the total action there may be other terms
that are second order in $\chi^{lm}$. 

Next, we rewrite (\ref{eq;3D-12}) in a 
 useful form in order to analyze the quantum correction
due to the coupling of the dilaton to the 
kinetic term of $\chi$. After the
some calculations, we obtain 
 
\begin{equation}
I_{\chi}=-\frac{1}{2}\int d^4x \sqrt{-\bar{g}^{(0)}}\,\chi_{lm}\left[
     \frac{1}{2}\Box \,{\rm e}^{-4\frac{\sigma}{\sigma_0}}
  -{\rm e}^{-4\frac{\sigma}{\sigma_0}}\Box+M^2_{\chi}\right]\chi_{lm}\,.
       \label{eq;3D-13}
\end{equation}
As in the case of the scalar field, we define the zeta function 
in order to compute the 1-loop quantum correction as 
%
\begin{equation}
\zeta_{\chi}(s)
         =\sum^{\infty}_{l=2}\sum^{\infty}_{l'=0}D_{\chi}(l)d_{\chi}(l')
        \left[\frac{\lambda_{\chi}(l')}{a^2}\,
         {\rm e}^{-4\frac{\sigma}{\sigma_0}}
         +{\rm e}^{-13\frac{\sigma}{\sigma_0}}\Lambda_{\chi}^2\right]\,,
       \label{eq;3D-14}
\end{equation}
where $D_{\chi}(l)$ and $d_{\chi}(l')$ are given by 
%
\begin{eqnarray}
D_{\chi}(l)&=&\frac{40(l-1)(l+2)(l+3)^2(l+4)(l+7)}{7!}\,,\\
d_{\chi}(l')&=&\frac{1}{6}(l+1)(l+4)(2l+3)\,, 
\end{eqnarray}
and $\lambda_{\chi}(l')$ and $\Lambda_{\chi}(l)$ are expressed as 
%
\begin{eqnarray}
\Lambda^2_{\chi}(l)&=&\frac{l(l+6)+30}{b_0^2}
         +{\rm e}^{9\frac{\sigma}{\sigma_0}}
          \left\{\frac{12}{a^2}
         -2\Lambda \,{\rm e}^{-7\frac{\sigma}{\sigma_0}}\right\}\,,\\
\lambda^2_{\chi}(l')&=&\frac{1}{2}l(l+3)\,. 
\end{eqnarray}
Using the results given in Appendix \ref{sec:scalar}, we can compute 
the values of
$\zeta_{\chi}(s)$ and $\zeta_{\chi}'(s)$ at $s=0$. 
Under the assumption that $a\gg b_0$, most integrals in 
$\zeta_{\chi}(s)$
 can be calculated analytically, or at least they can be 
put into simple forms, that can be computed numerically.
  After tedious calculations, we obtain the expression 
%
\begin{equation}
V_{\rm eff}(\sigma)=\frac{\bar{\Lambda}}{\kappa^2}
           \,{\rm e}^{-7\frac{\sigma}{\sigma_0}}
 	   -\frac{21}{\kappa^2b_{0}^2}
            {\rm e}^{-9\frac{\sigma}{\sigma_0}}
           +\frac{5.93731\times 10^{-5}}{b_0^4}
             \,{\rm e}^{-18\frac{\sigma}{\sigma_0}}\,.
       \label{eq;3D-17}
\end{equation}
 
\subsection{Quantum correction in $AdS_4\times S^7$}
\label{subsec:AdS4}

As seen in the last section, the dilaton potential 
energy is negative at the minimum.
Then, we have to calculate the quantum correction 
in the $AdS_4\times S^7$ background and derive the effective potential.

Because the procedure to obtain the effective potential in
 $AdS_4\times S^7$, 
is almost the same as that in $dS_4\times S^7$,
 except for the zeta functions, we do not repeat it here.
Instead, showing how to regularize the zeta functions in $AdS_4\times S^7$
 in Appendix \ref{sec:zads} or Ref.\,39),
we just summarize our results as follows.

The effective potentials for bosonic fields
 (the scalar field $\bar{\phi}$,
the $U(1)$ gauge field $A_M$, and scalar mode of gravitational field
 $h^{lm}_{\rm (T)}$) are given by 
\begin{eqnarray}
V_{\rm eff}(\sigma)=\frac{\bar{\Lambda}}{\kappa^2}
          \,{\rm e}^{-7\frac{\sigma}{\sigma_0}}
 	   -\frac{10}{\kappa^2b_{0}^2}
        \,{\rm e}^{-9\frac{\sigma}{\sigma_0}}   
        +\frac{D_{\rm c}}{b_0^4}\,{\rm e}^{-18\frac{\sigma}{\sigma_0}}\,,
\label{QC_AdS}
\end{eqnarray}
where $D_{\rm c}$ is given by 
%
\begin{eqnarray}
D_{\rm c}
&& = 6.03329\times10^{-3} ~~~{\rm for~a~scalar~field}\,,\nonumber \\
&& =-7.21715\times10^{-6} ~~~{\rm for~a~Dirac~spinor~field}\,,\nonumber \\
&& = 9.97116\times10^{-3} ~~~{\rm for~a~gauge~field}\,,\nonumber \\
&& = 8.21713\times10^{-5} ~~~{\rm for~a~gravitational~field}\,.
\end{eqnarray}
 
\subsection{Dilaton dynamics and stabilization of $S^7$}
We expect that the effective potential can be used to judge the
stability of the solution with respect to the spacetime independent
 dilaton or contractions of the compact manifold.   
If the extra dimensions ($S^7$) are stable, the dilaton effective 
potential must possess a minimum or at least a local minimum.  From 
(\ref{eq;potential}), (\ref{eq;fpotential}), (\ref{eq;vpotential}) 
and (\ref{eq;3D-17}), 
the effective potential in $dS_4\times S^7$ is given by 
%
\begin{eqnarray}
V_{\rm eff}(\sigma)=\frac{\bar{\Lambda}}{\kappa^2}\,
          {\rm e}^{-{7}\frac{\sigma}{\sigma_0}}
 	   -\frac{21}{\kappa^2b_{0}^2}\,
        {\rm e}^{-9\frac{\sigma}{\sigma_0}}   
         +\frac{C}{b_0^4}\,{\rm e}^{-{18}\frac{\sigma}{\sigma_0}}\,,
\label{eq;dS}
\end{eqnarray}
where $C$ is given by 
%
\begin{eqnarray}
C=N_{\rm S} C_{\rm S}+N_{\rm F}C_{\rm F}+N_{\rm V} C_{\rm V}+C_{\chi}\,, 
\end{eqnarray}
with 
$N_{\rm S}, N_{\rm F}$, 
and $N_{\rm V}$ the numbers of the scalar, Dirac spinor
 and vector field.

\begin{figure}[b]
            \epsfxsize = 11 cm
            \epsfysize = 9 cm
            \centerline{\epsfbox{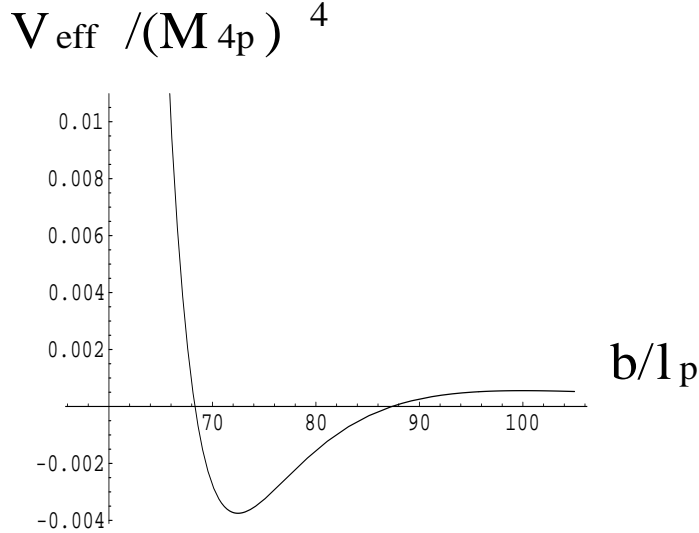}}
\caption{The dilaton effective potential in Eq.\,(\ref{eq;dS}). Note that the local minimum of the dilaton potential 
is $b= 72.4691 l_p$. Here $M_{4p}$ denotes the four-dimensional Planck mass.}
        \label{fig:1}
\end{figure}
As shown in Fig.\,\ref{fig:1}, we find the a minimum of 
the potential at $72.4691 l_p$ if we take 
$N_{\rm S}=N_{\rm V}=N_{\rm F}=10^5$  
$\Lambda= 2.62723\times 10^{-3}M^2_p$, 
and $b_0= 100 l_p$,  where $M_p$ and $l_p$
denote the Planck mass and Planck scale, respectively.
We have also calculated the back reaction due to changing the
background geometry.
 Because the effective potential is zero at $b=0.873448b_0$, the
background geometry is the four-dimensional Minkowski spacetime 
$M_4$. Thus we need to calculate the quantum correction in the 
$M_4$ fixed background.
 However, the effective potential calculated in $M_4$ is also 
negative at this minimum point. That is, 
the effective potential immediately becomes 
 negative. Then, we have to calculate the quantum
correction in the 
negative cosmological constant. We obtain the
effective potential derived in the spacetime with negative cosmological
constant,\cite{uz2} which is the
potential minimum at $b= 0.669486b_0$. Note that the geometry is 
$AdS_4$ at the local minimum of the potential.   

 Finally, we discuss the difference between $M_4\times S^7$ and
 $(A)dS_4\times S^7$. The effective potential of the scalar and spinor
 fields in $M_4\times S^d$ was calculated by Candelas and Weinberg
 \cite{can} to be 
%
\begin{eqnarray}
V_{\rm eff}(\sigma)=U(\sigma)+\frac{f_{\rm M}}{b_0^4}
  {\rm e}^{-2(d+2)\frac{\sigma}{\sigma_0}}\,, 
\end{eqnarray}
where $f_{\rm M}$ is a numerical coefficient.
Similarly, the effective potential in $dS_4\times S^d$ is given
by   
%
\begin{eqnarray}
V_{\rm eff}(\sigma)=U(\sigma)+\frac{f_{\rm d}}
 {b_0^4}{\rm e}^{-2(d+2)\frac{\sigma}
  {\sigma_0}}\,. 
\end{eqnarray}
Thus, we see that
 there is no difference between the two cases. This is the reason why the 
condition $a\gg b_0$ in $dS_4\times S^d$ is assumed 
in the calculation of the effective potential. The full result is the
given by   
%
\begin{eqnarray}
V_{\rm eff}(\sigma)=U(\sigma)+\frac{f_1}{b_0^4}\,{\rm e}^{-2(d+2)\frac{\sigma}
  {\sigma_0}}+\frac{f_2}{a^2b_0^2}\,
   {\rm e}^{-(d+2)\frac{\sigma}{\sigma_0}}\,. 
    \label{eq;rep}
\end{eqnarray}
Because we make the assumption
%
\begin{eqnarray}
\left(\frac{b_0}{a}\right)10^{-8}\ll 1 \hspace{0.5cm} \longrightarrow 
\hspace{0.5cm} \frac{f_2}{a^2 b_0^2}\ll \frac{f_1}{b_0^4}\,,\hspace{1cm}
(f_1\sim f_2)
\end{eqnarray}
the effective potential in $dS_4\times S^d$ is given by
Eq.\,(\ref{eq;rep}). In the Minkowski limit $a\rightarrow\infty$, the
effective potential in the $dS_4\times S^d$ is equal to the form of the
effective potential in $M_4\times S^7$. 
The effective action and effective potential depend
 on the parameter $b_0$ (the initial value of $b$). 
This is the reason why we have calculated the
 quantum correction of the various fields in the $dS_4\times S^7$ 
fixed background. The four-dimensional spacetime for $b=b_0$ becomes 
$dS_4$.
Then, the quantum correction for $b\ne b_0$ generally cannot be
calculated, because it is difficult to solve the field equation 
in the unknown spacetime $\hat{g}_{\mu\nu}$. 
 By the rescaling the matter field, however, 
the calculation of the effective potential can be performed 
with the fixed background $dS_4\times S^7$.
Therefore, the final results depend on the parameter $b_0$. 

%% file: sec4.tex
\section{Dilaton stabilization with eleven-dimensional supergravity}
	\label{sec:sugra}
In this section, we consider dilaton stabilization in 
an eleven-dimensional supergravity model. 
This model is considered to be a possible theory to provide
 a low energy description of M-theory. 
We consider the model in which the background geometry
 is a product of
 $dS_4$ and $S^7$ and assume that the radius of $S^7$ is
 microscopically small. 
Although the supersymmetry is broken in $dS_4$, 
this causes no problem from the cosmological point of view.
   
We consider the bosonic part of the action for $N=1$ supergravity in
eleven-dimensional spacetime, 
%
\begin{eqnarray}
I&=&\int d^{11}x\sqrt{-\bar{g}}\left\{\frac{1}{2\:\bar{\kappa}^2}\bar{R}
       -\frac{1}{48}\:F_{MNPQ}\:F^{MNPQ}\right.\nonumber\\
 &+&\left.
     \frac{\sqrt{2}}{6\cdot(4!)^2}\epsilon^{M_1\cdots M_{11}}
    F_{M_1\cdots M_4}\:F_{M_5\cdots M_8}\:A_{M_9\cdots M_{11}}\right\}\,,
           \label{eq;4-1}
\end{eqnarray}
where $\epsilon_{M_1\cdots M_{11}}$ is the Levi-Civita symbol.
The background line element is assumed to take the 
form given in (\ref{eq;2-2}).
We assume that the gauge field strength has the Freund-Rubin form, 
proportional to Levi-Civita tensor in four-dimensional spacetime: 
%
\begin{eqnarray}
F_{M_1\cdots M_4}&=&\left\{ 
\begin{array}{ll}
 \left(f/\sqrt{\Omega}\right)\epsilon_{M_1\cdots M_4}\,,& 
M_1=\mu_1, \cdots, M_4=\mu_4\\
 0\,, & {\rm otherwise}
\end{array}
\right.
 \\
F_{\mu_1\mu_2\mu_3\mu_4}\:F^{\mu_1\mu_2\mu_3\mu_4}&\equiv&96\Lambda\,.
           \label{eq;4-3}
\end{eqnarray}
Here, $f$ is the constant and 
 $(\mu_1, \cdots, \mu_4)=0,\cdots,3$. 
Then, the action is given by 
%
\begin{equation}
I=\int d^{11}x\sqrt{-\bar{g}}\left(\frac{1}{2\bar{\kappa}^2}\bar{R}
       -2\Lambda\right)\,,
\end{equation}
which has the same form as (\ref{eq;2-1}). Now 
we consider the gravitational perturbation (\ref{eqn;2}) and quantum
fluctuations of the scalar mode of $h_{MN}$.
The procedure for calculation of the quantum correction is 
the same as (D) in $\S$\ref{subsec;EP}. 
Therefore the dilaton potential is given by 
(\ref{eq;3D-17}). The local minimum for the dilaton potential 
is located at $b\simeq 0.03b_0$.  Although 
the potential energy of the dilaton is initially positive, the dilaton 
settle down to the minimum of its potential, which is a negative 
value.


%% file: sec5.tex
\section{Conclusion}
	\label{sec:conclusion}
In this paper, we have investigated the stabilization of the extra dimension 
in a higher-dimensional theory. 
In particular, we have focused on the eleven-dimensional Kaluza-Klein and
supergravity. 
As the source of the stabilization, we have
considered a quantum effect associated with 
background matter field.  We have studied the stability for the  
compactification, which is the product spacetime $dS_4\times S^7$.
We have considered the case that the scale of  $dS_4$ is far larger 
than that of the extra dimension and demonstrated the 
compactification model,
 in which the dilaton potential has a local minimum
with respect to the quantum correction of the matter fields.  
The part coming from the quantum effects in the dilaton
 effective potential is dominant for $b\rightarrow l_p$, while the
cosmological constant in it is dominant for large $b$.       
The dilaton field in our model does not evolve forever
 but, rather, settles down to its potential minimum. 
Consequently, the extra dimension is asymptotically almost
 time independent in $dS_4$ spacetime, and its scale is 30 $l_p$. 
Furthermore, we have found that the dilaton potential energy
 is negative at its minimum. Because the dilaton potential energy 
 is the effective cosmological constant in four dimensions,
  the $dS_4$ spacetime is transformed to a spacetime with
 negative cosmological constant after the dilaton has settled down
 at the local minimum of the dilaton potential. 
Our model describes the situation in which a  
geometrical change has occurred over the whole spacetime, because the
bulk cosmological constant $\Lambda$ and the dilaton field are
homogeneous. 
This mechanism is applicable to a model in which the
background geometry changes from $dS_5$ to $AdS_5$. We will 
discuss the dynamics of this change in another paper.\cite{uz2}

In the very early universe, the scale of $dS_4$ may have been
on the same orders as that of $S^d$\,. Therefore, the dynamics of $b$ and
the behavior of its effective potential are cosmologically very
interesting. However, the integral in $\zeta(s)$ cannot be naively 
calculated in the case $a\simeq b_0$.  This point will be considered 
in a future work. 
\\
\\

\section*{Acknowledgements}
We would like to thank K. Maeda, T. Kubota, O. Yasuda and N. Sakai 
for valuable comments.
We would also like to thank M. Sakagami and J. Soda for 
continuing encouragement and useful comments.

%% file: appenA.tex
\section{Zeta Function Regularization for a Scalar Field}
         \label{sec:scalar}
\begin{figure}[t]
            \epsfxsize = 10 cm
            \epsfysize = 8 cm
            \centerline{\epsfbox{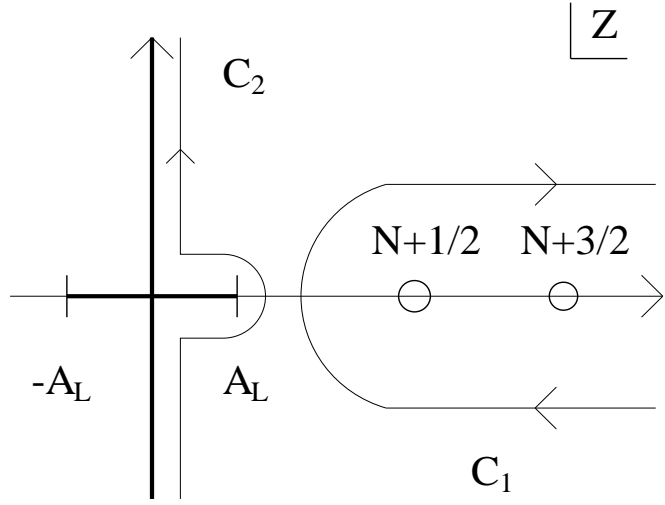}}
        \caption{The contour $C_1$ in Eq.\,(\ref{eq;a-9}) is replaced by
 the contour $C_2$. Note that the contour $C_2$ avoids the branch point at
 $z=\pm A_{\rm L}$.}
        \label{fig:s1}
\end{figure}

In this appendix, we present a method to regularize the zeta function
for an odd total number of dimensions that has the product geometry 
 $dS_n\times S^d$. As mentioned in $\S$\ref{subsec;EP}, 
this geometry is identical to
$S^n\times S^d$ in the Euclidean spacetime after performing the Wick
rotation in $dS^n$. 
It can easily seen that this method is applicable for an  
even number of dimensions.\footnote{Strictly speaking, in an 
even number of
dimensions, a conformal anomaly arises, and the arguments become more 
 complicated.} These formulations are similar to those  
proposed by Kikkawa et al.\cite{ki}
 We employ the method and notation used 
by them. First, we define a generalized zeta function for the
scalar fields in $S^n\times S^d$ as
%
\begin{eqnarray}
\zeta(s)&=&\sum^{\infty}_{l'=0}\sum^{\infty}_{l=0}
         \frac{(l'+n-2)!\;(l+d-2)!}{(n-1)!\;(d-1)!}
         \frac{(2l'+n-1)\;(2l+d-1)}{l'!\hspace{0.1cm}l!}
            \nonumber\\
        & & \hspace{0.5cm}
            \times\left\{\frac{l'(l'+n-1)}{a^2}
            +\frac{l(l+d-1)}{b_0^2}
             \left(\frac{b_0}{b}\right)^{d+2}\right\}^{-s}\,,
       \label{eq;a-1}
\end{eqnarray}
where we assume that $n$ is an even number 
and $d$ is an odd number.
Using $D=(d-1)/2$ and $N=n/2$ instead of $d$ and $n$ and the 
running variables
$L=l+D$ and $L'=l'+N$, we rewrite (\ref{eq;a-1}) as  
%
\begin{equation}
\zeta(s)=\sum^{\infty}_{L'=N}D_n\left(L'-\frac{1}{2}\right)
         \sum^{\infty}_{L=D}D_d(L)
         \left\{\frac{\Lambda_n\left(L'-\frac{1}{2}\right)}{a^2}
         +\frac{\Lambda_d(L)}{b_0^2}
         \left(\frac{b_0}{b}\right)^{d+2}\right\}^{-s}\,, 
         \label{eq;a-2}
\end{equation}
where the variables are given by
%
\begin{eqnarray}
D_n\left(L'-\frac{1}{2}\right)&=&
             \frac{2L'-1}{(2N-1)!}\left\{\left(L'-\frac{1}{2}\right)^2
             -\left(N-\frac{3}{2}\right)^2\right\}
             \cdots\left\{\left(L'-\frac{1}{2}\right)^2
             -\left(\frac{1}{2}\right)^2\right\}\,,\nonumber\\
D_d(L)&=&\frac{2L^2}{(2D)!}\left\{L^2-(D-1)^2\right\}
            \cdots\left\{L^2-1\right\},\nonumber\\
\Lambda_n\left(L'-\frac{1}{2}\right)&=&\left(L'-\frac{1}{2}\right)^2
             -\left(N-\frac{1}{2}\right)^2\,,\nonumber\\
\Lambda_d(L)&=&L^2-D^2\,.
        \label{eq;a-3}
\end{eqnarray}
Now, we replace the infinite mode sum over $L'$ by complex integration.
The generalized zeta function (\ref{eq;a-2}) is then 
%
\begin{eqnarray}
\zeta(s)&=&(a)^{2s}\sum^{\infty}_{L=D}D_d(L)\frac{-i}{2}\int_{C_1}dz
         \tan(\pi z)D_n(z)\nonumber\\
        & & \times\left\{z^2+\left(\frac{a}{b_0}\right)^2
         \left(\frac{b_0}{b}\right)^{d+2}\Lambda_d(L)
         -\left(N-\frac{1}{2}\right)^2\right\}^{-s}\,,
       \label{eq;a-4}
\end{eqnarray}
where the contour $C_1$ in the complex plane is showed 
in Fig.\,\ref{fig:s1}.
In order to avoid the singularity of $z$ in the integrand, it is
convenient to divide the sum over $L$ into two parts,
%
\begin{eqnarray}
\zeta(s)=Z(s)+W(s)\,,
       \label{eq;a-5}
\end{eqnarray}
where
%
\begin{eqnarray}
Z(s)&=&(a)^{2s}\sum^{L_0}_{L=D}D_d(L)\frac{-i}{2}\int_{C_1}dz
         \tan(\pi z)D_n(z)\left(z^2-A_{\rm L}^2\right)^{-s}\,,\nonumber\\
W(s)&=&(a)^{2s}\sum^{\infty}_{L=L_0+1}D_d(L)\frac{-i}{2}\int_{C_1}dz
         \tan(\pi z)D_n(z)\left(z^2+B_{\rm L}^2\right)^{-s}\,,
       \label{eq;a-6}
\end{eqnarray}
with
%
\begin{eqnarray}
A_{\rm L}^2&=&\left(N-\frac{1}{2}\right)^2
      -\left(\frac{a}{b_0}\right)^2\left(\frac{b_0}{b}\right)^{d+2}
      \Lambda_d(L)\,,\nonumber\\
B_{\rm L}^2&=&-A_L^2\,.
       \label{eq;a-7}
\end{eqnarray}
Here, $A_{\rm L}^2$ and $B_{\rm L}^2$
 are positive for $D\le L<L_0$ and $L_0\le L$,
 respectively. $L_0$ is defined as the largest integer smaller than
 or equal to 
%
\begin{equation}
\sqrt{D^2+\left(\frac{b_0}{a}\right)^2\left(\frac{b}{b_0}\right)^{d+2}
\left(N-\frac{1}{2}\right)^2}\,.
       \label{eq;a-8}
\end{equation}
We consider the function $Z(s)$, which is a sum over the $L$ covers from 
$D$ to $L_0$. There are branch points at $z=\pm A_{\rm L}$. 
Now, we replace the
contour $C_1$ so as to move in a direction
 parallel to the imaginary axis in order
to deal with the poles of $\tan(\pi z)$ (see Fig.\,\ref{fig:s1}). 
\noindent
Because the displacing contour 
$C_2$ is defined as running above the cut associated with
 $z=\pm A_{\rm L}$, the result for $Z(s)$ is given by  
%
\begin{eqnarray}
Z(s)&=&(a)^{2s}\sum^{L_0}_{L=D}D_d(L)\nonumber\\
    & &\times\left\{\frac{1}{2}
      \left({\rm e}^{-i\pi s}
     +{\rm e}^{i\pi s}\right)\int^{\infty}_0dx\;i\:D_n(ix)
      \left(x^2+A_{\rm L}^2\right)^{-s}\tanh(\pi x)\right.\nonumber\\
    & &-\frac{i}{2}\,e^{i\pi s}\int^{A_{\rm L}}_0
      dx\:\tan\pi(x-i\epsilon)D_n(x)\left(A_{\rm L}^2-x^2\right)^{-s}
      \nonumber\\
    & &\left.+\frac{i}{2}\,e^{-i\pi s}\int^{A_{\rm L}}_0
      dx\:\tan\pi(x+i\epsilon)D_n(x)\left(A_{\rm L}^2-x^2\right)^{-s}
      \right\}\,,
       \label{eq;a-9}
\end{eqnarray}
where $D_n(ix)$ is the polynomial with coefficients $r_{Nk}$:
%
\begin{eqnarray}
D_n(ix)&=&i(-1)^{N-1}\frac{2x}{(2N-1)!}
        \left\{x^2+\left(N-\frac{3}{2}\right)^2\right\}\cdots
        \left\{x^2+\left(\frac{1}{2}\right)^2\right\}\nonumber\\
       &\equiv&i(-1)^{N-1}\sum^{N-1}_{k=0}\;r_{Nk}\;x^{2k+1}\,.
       \label{eq;a-10}
\end{eqnarray}
The first term in Eq.\,(\ref{eq;a-9}) comes from the integral 
along the imaginary
axis. The second and third terms in Eq.\,(\ref{eq;a-9}) 
are the contributions from the
contour along the cut on the real axis.
Since we use the relation 
%
\begin{equation}
\tanh(\pi x)=1-\frac{2}{{\rm e}^{2\pi x}+1}\,,
       \label{eq;a-11}
\end{equation}
and substitute it into the first term of (\ref{eq;a-9}), 
the function $Z(s)$ is   
%
\begin{eqnarray}
Z(s)&=&-(a)^{2s}\sum^{L_0}_{L=D}D_d(L)\nonumber\\
    & &\times\left\{\cos(\pi s)
      \frac{1}{\Gamma(s)}(-1)^{N-1}\frac{1}{2}\sum^{N-1}_{k=0}f_{Nk}
      \left(A_{\rm L}^2\right)^{k+1-s}\Gamma(k+1)\Gamma(s-k-1)\right.
      \nonumber\\
    & &+\cos(\pi s)\int^{\infty}_0 dx\;i\:D_n(ix)
       \left(x^2+A_{\rm L}^2\right)^{-s}
       \frac{2}{{\rm e}^{2\pi x}-1}
      \nonumber\\
    & &\left.+\sin(\pi s)\int^{A_{\rm L}}_0 dx D_n(x)
        \left(A_{\rm L}^2-x^2\right)^{-s}\tan(\pi x)\right\}\,. 
         \label{eq;a-12}
\end{eqnarray}
We note that third term in Eq.\,(\ref{eq;a-12}) 
does not contribute to $\zeta(0)$.

\begin{figure}[t]
            \epsfxsize = 10 cm
            \epsfysize = 8 cm
            \centerline{\epsfbox{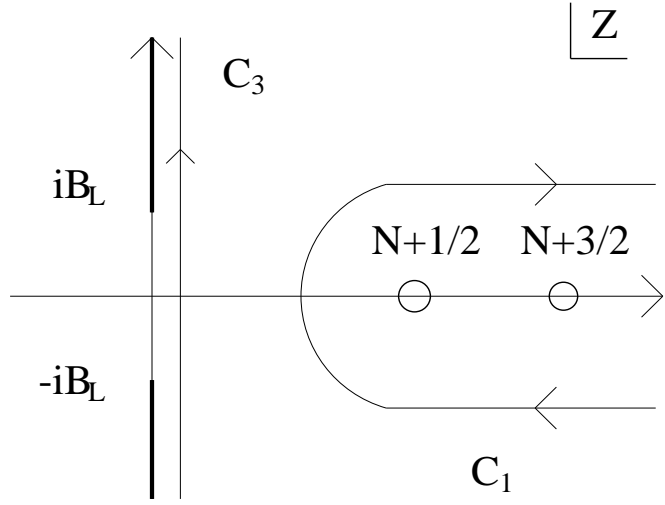}}
        \caption{The contour $C_1$ in (\ref{eq;a-12}) replaced by
 the contour $C_3$. The path of $C_3$ runs parallel to the
 imaginary axis.}
        \label{fig:s2}
\end{figure}

\begin{figure}[t]
            \epsfxsize = 9 cm
            \epsfysize = 7.5 cm
            \centerline{\epsfbox{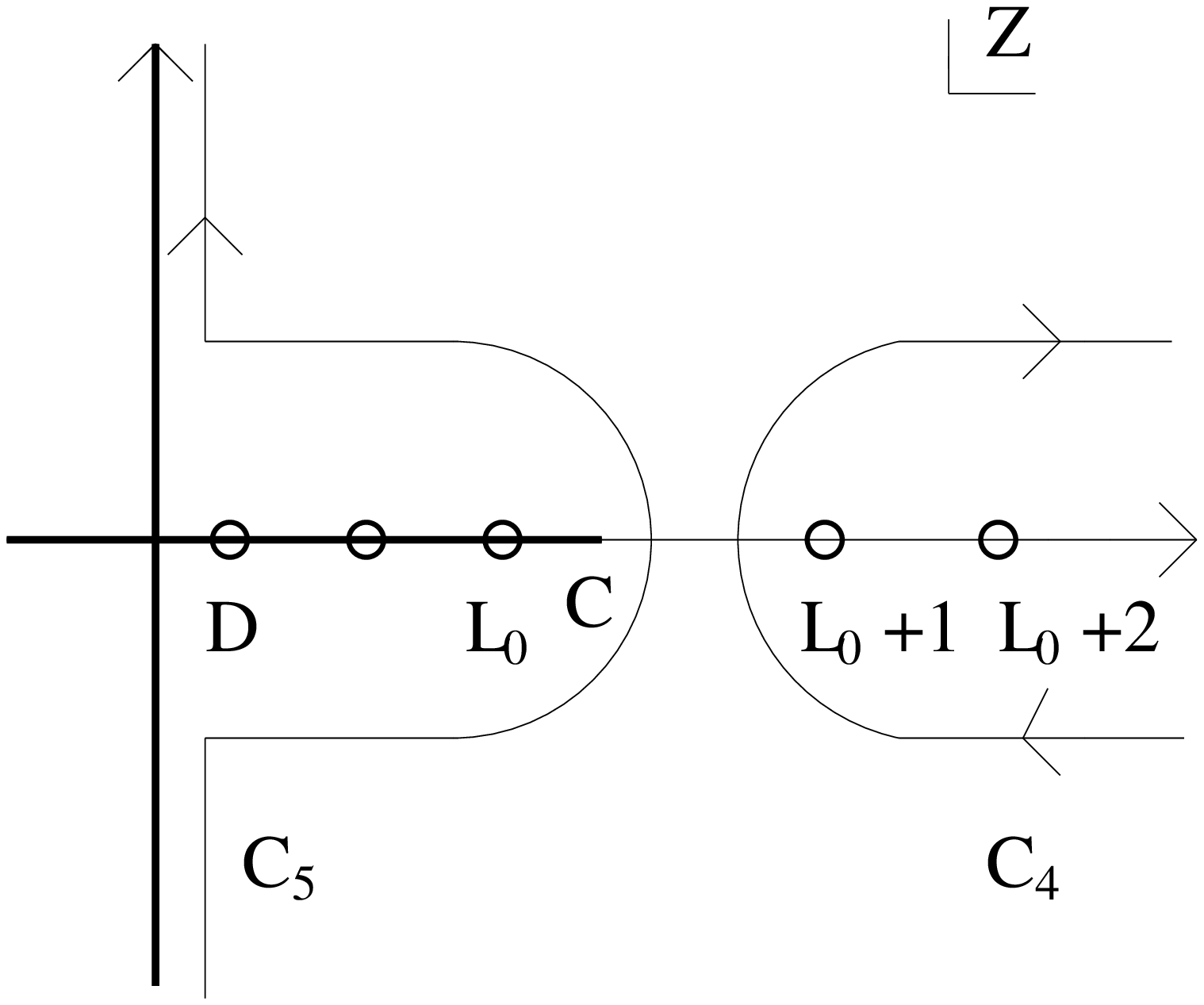}}
        \caption{The contour $C_4$ in (\ref{eq;a-16}) replaced 
by the contour $C_5$. The pole of $\cos(\pi z)$ exists at $z=L_0$,
 which is a smaller value than that at the cut of the branch point. }
        \label{fig:s3}
\end{figure}

Next, we consider the function $W(s)$. The branch points in the integrand 
are on the imaginary axis at $z=\pm iB_{\rm L}$.
Replacing the contour $C_1$ by $C_3$, we obtain
%
\begin{eqnarray}
W(s)&=&(a)^{2s}\sum^{\infty}_{L=L_0+1}D_d(L)
     \int^{\infty}_0 dx\;iD_n(ix)\tanh(\pi x)
     \left(B_{\rm L}^2-x^2\right)^{-s}
      \nonumber\\
    &=&(a)^{2s}\sum^{\infty}_{L=L_0+1}D_d(L)\left[
      \frac{1}{\Gamma(s)}(-1)^N\frac{1}{2}\sum^{N-1}_{p=0}r_{Np}
      \left(B_{\rm L}^2\right)^{p+1-s}\right.\nonumber\\
    & &\times
     \left\{\frac{\Gamma(s-p-1)\Gamma(-s+1)}{\Gamma(-p)}\cos(\pi s)
      +\frac{\Gamma(p+1)\Gamma(-s+1)}{\Gamma(2+p-s)}\right\}
           \nonumber\\
    & &-\cos(\pi s)\int^{\infty}_{B_{\rm L}} dx\:i\:D_n(ix)
       \left(x^2-B_{\rm L}^2\right)^{-s}
       \frac{2}{{\rm e}^{2\pi x}+1}
           \nonumber\\
    & &\left.-\int^{B_{\rm L}}_0 dx\:i\:D_n(ix)
       \left(B_{\rm L}^2-x^2\right)^{-s}
       \frac{2}{{\rm e}^{2\pi x}+1}\right]\,,\nonumber\\
         \label{eq;a-13}
\end{eqnarray}
where the contour $C_3$ in the complex plane is showed in
 Fig.\,\ref{fig:s2}.
\noindent
Note that  
the function $W(s)$ includes an infinite sum over $L$. As far as the
second term of Eq. (\ref{eq;a-13}) 
is concerned, the infinite sum over $L$ is
convergent, due to the fact that 
integral decreases exponentially as $L$ goes to
infinity. For the first term in Eq. (\ref{eq;a-13}), 
 however, we must regularize the summation  
%
\begin{equation}
\sum^{\infty}_{L=L_0+1}D_d(L)\left(B_{\rm L}^2\right)^{p+1-s}
             =\left\{\left(\frac{a}{b_0}\right)^{2}
              \left(\frac{b_0}{b}\right)^{d+2}\right\}^{p-s+1}
         \sum^{\infty}_{L=L_0+1}D_d(L)\left(L^2-C^2\right)^{-\beta}\,,
           \label{eq;a-14}
\end{equation}
with 
%
\begin{equation}
C^2=D^2+\left(\frac{b_0}{a}\right)^2\left(\frac{b}{b_0}\right)^{d+2}
    \left(N-\frac{1}{2}\right)^2\,,
           \label{eq;a-15}
\end{equation}
where $\beta=s-p-1$. 
Again replacing the infinite sum with a complex integral, we obtain  
%
\begin{equation}
S\equiv\sum^{\infty}_{L=L_0+1}D_d(L)\left(L^2-C^2\right)^{-\beta}
 =\frac{-1}{2i}\int_{C_4}dz D_d(z)\left(z^2-C^2\right)^{-\beta}
    \cot(\pi z)\,,
           \label{eq;a-16}
\end{equation}
where the sum is taken over half integers and the contour $C_4$ is
that depicted in Fig. \ref{fig:s3}.
\noindent
 By replacing the contour $C_4$ by $C_5$, we obtain 
%
\begin{eqnarray}
S&=&-\cos(\pi \beta)\sum^{L_0}_{L=D}D_d(L)\left(C^2-L^2\right)^{-\beta}
    \nonumber\\
 & & +\sin(\pi \beta)\left\{\int^{\infty}_{0}dx\;
    D_d(ix)\left(x^2+C^2\right)^{-\beta}
    \left(\frac{2}{{\rm e}^{2\pi x}-1}\right)\right.\nonumber\\
 & &+\int^{C}_{0}dx\;D_d(x)\left(C^2-x^2\right)^{-\beta}
    \cot(\pi x)\nonumber\\
 & &\left.+(-1)^{D}\frac{1}{2}
    \sum^{D-1}_{q=0}\;r_{Dq}\;\left(C^2\right)^{q-\beta+\frac{3}{2}}
    \frac{\Gamma\left(q+\frac{3}{2}\right)
    \Gamma\left(\beta-q-\frac{3}{2}\right)}{\Gamma\left(\beta\right)}
   \right\}\,,
           \label{eq;a-17}
      \nonumber\\
\end{eqnarray}
where the function $D_d(ix)$ is defined as 
%
\begin{eqnarray}
D_d(ix)&=&(-1)^{D}\frac{2x^2}{(2D)!}
        \left\{x^2+\left(D-1\right)^2\right\}\cdots
        \left\{x^2+1^2\right\}\nonumber\\
       &\equiv&(-1)^{D}\sum^{D-1}_{q=0}\;r_{Dq}\;x^{2q+2}\,,
             \label{eq;a-18}
\end{eqnarray}
$P$ stands for the principal value integral, and 
the coefficients $j_{Dq}$ in $D_{d}$ are defined in Eq.\,(\ref{eq;a-18}).
Substituting (\ref{eq;a-18}) into the first term of 
Eq.\,(\ref{eq;a-13}), we obtain an expression for $W(s)$. 
Finally, the generalized $\zeta$ function is found to be  
%
\begin{eqnarray}
\zeta(s)&=&(a)^{2s}\sum^{L_0}_{L=D}D_d(L)\nonumber\\
        & &\times\left[\cos(\pi s)\left\{
      \frac{1}{\Gamma(s)}(-1)^N\frac{1}{2}\sum^{N-1}_{k=0}r_{Nk}
      \left(A_{\rm L}^2\right)^{k-s+1}\Gamma(k+1)\;\Gamma(s-k-1) 
        \right.\right.\nonumber\\
    & &\left.
       -\int^{\infty}_0 dx\;i\:D_n(ix)\left(x^2+A_{\rm L}^2\right)^{-s}
       \frac{2}{{\rm e}^{2\pi x}+1}\right\}
          \nonumber\\
    & &\left.-\sin(\pi s)
       \int^{A_{\rm L}}_0 dx D(x)
       \left(A_{\rm L}^2-x^2\right)^{-s}\tan(\pi x)
      \right]\nonumber\\
    & &-(a)^{2s}\sum^{\infty}_{L=L_0+1}D_d(L)\;\left\{
       \int^{\infty}_{B_{\rm L}}
       dx\;i\:D_n(ix)\left(x^2-B_{\rm L}^2\right)^{-s} 
       \frac{2}{{\rm e}^{2\pi x}+1}\cos(\pi s)\right.\nonumber\\
    & &\left.
       +\int^{B_{\rm L}}_0 dx\;i\:D_n(ix)
       \left(B_{\rm L}^2-x^2\right)^{-s} 
       \frac{2}{{\rm e}^{2\pi x}+1}\right\}\nonumber\\
    & &+(a)^{2s}(-1)^N\frac{1}{2}
       \sum^{N-1}_{p=0}r_{Np}\left\{\left(\frac{a}{b_0}\right)^2
       \left(\frac{b_0}{b}\right)^{d+2}\right\}^{p-s+1}
       \nonumber\\
    & &\times
      \left\{\frac{\Gamma(s-p-1)\Gamma(-s+1)}{\Gamma(-p)}\cos(\pi s)
       +\frac{\Gamma(p+1)\Gamma(-s+1)}{\Gamma(2+p-s)}
       \right\}\times S\,.   
           \label{eq;a-19}
\end{eqnarray}

%% file: appenB.tex
\section{Zeta Function Regularization for a Spinor Field}
         \label{sec:spinor}
\begin{figure}[t]
            \epsfxsize = 10 cm
            \epsfysize = 8 cm
            \centerline{\epsfbox{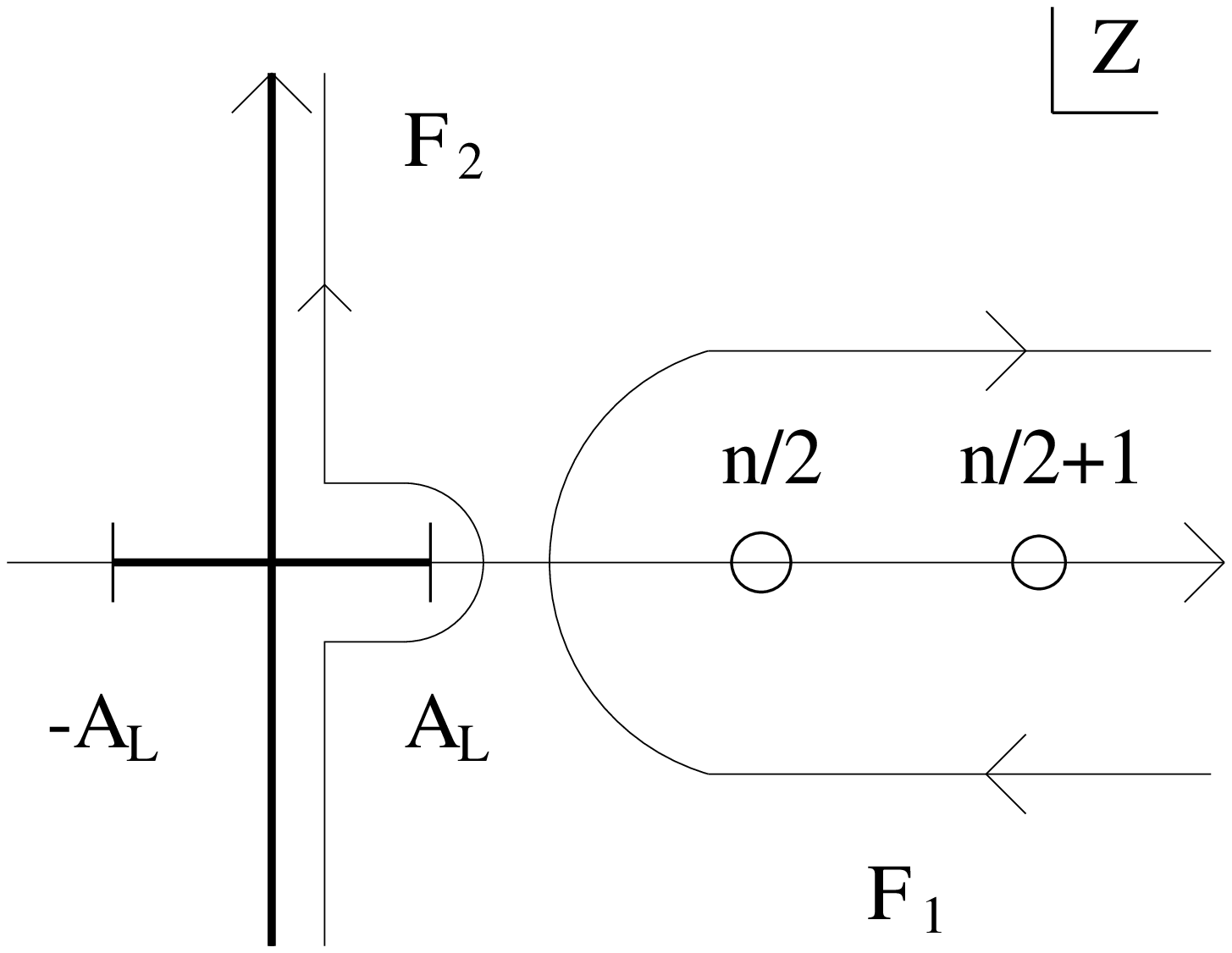}}
\caption{The contour $F_1$ in Eq.\,(\ref{eq;b-4}) is replaced by the
 contour $F_2$. Note that the contour $F_2$ avoids the branch points at
 $z=\pm A_{\rm L}$\,.}
        \label{fig:f1}
\end{figure}

In this appendix, we present a method to regularize the zeta function
for the Dirac spinor field in an odd total number of dimensions.
 It can easily be seen that this
method is able to apply the case of an even number of dimensions. 
 We define the generalized zeta function for the
Dirac spinor fields in $S^n\times S^d$ as
%
\begin{equation}
\zeta(s)=\sum^{\infty}_{l'=0}\sum^{\infty}_{l=0}
         \frac{(l'+n-1)!\;(l+d-1)!}{(n-1)!\;(d-1)!\;l'!\;l!}
      \left\{\frac{\left(l'+\frac{n}{2}\right)^2-\frac{n(n-1)}{4}}{a^2}
            +\frac{\left(l+\frac{d}{2}\right)^2}{b^2}
            \right\}^{-s}\,,
           \label{eq;b-1}
\end{equation}
where $n$ is an even number and $d$ is an odd number. 
Using the running variables
$L=l+d/2$ and $L'=l'+n/2$, we rewrite (\ref{eq;b-1}) as 
%
\begin{equation}
\zeta(s)=\sum^{\infty}_{L'=n/2}D_n\left(L'\right)
         \sum^{\infty}_{L=d/2}D_d(L)
         \left\{\frac{L'^2-\frac{n(n-1)}{4}}{a^2}
         +\frac{L^2}{b^2}\right\}^{-s}\,, 
           \label{eq;b-2}
\end{equation}
where the eigenvalues and degeneracies are now given by
%
\begin{eqnarray}
D_n\left(L'\right)&=&
             \frac{L'}{(n-1)!}\left\{L'^2-\left(\frac{n}{2}-1\right)^2
             \right\}
             \cdots\left(L'^2-1\right)\,,\nonumber\\
D_d(L)&=&\frac{1}{(d-1)!}\left\{L^2-\left(\frac{d}{2}-1\right)^2\right\}
            \cdots\left\{L^2-\left(\frac{1}{2}\right)^2\right\}\,.
           \label{eq;b-3}
\end{eqnarray}
Next, replacing the infinite mode sum over $L'$ by complex integration,  
the generalized zeta function is  
%
\begin{equation}
\zeta(s)=(a)^{2s}\sum^{\infty}_{L=d/2}D_d(L)\frac{-i}{2}\int_{F_1}dz
         \tan(\pi z)D_n(z)
         \left\{z^2+\left(\frac{a}{b}\right)^2\;L^2
         -\frac{n(n-1)}{4}\right\}^{-s}\,,
           \label{eq;b-4}
\end{equation}
where the contour $F_1$ in the complex plane is showed in 
Fig. \ref{fig:f1}.
\noindent
We divide the sum over $L$ into two parts, (\ref{eq;a-5}) and
 (\ref{eq;a-6}), with 
%
\begin{eqnarray}
A_{\rm L}^2&=&\frac{n(n-1)}{4}
      -\left(\frac{a}{b}\right)^2L^2\,,\nonumber\\
B_{\rm L}^2&=&-A_{\rm L}^2\,,
       \label{eq;b-5}
\end{eqnarray}
where $A_{\rm L}^2$ and $B_{\rm L}^2$ are positive 
for $D\le L<L_0$ and $L_0\le L$,
 respectively. The number
 $L_0$ is defined as the largest integer smaller than or equal to 
%
\begin{equation}
\sqrt{\frac{n(n-1)}{4}\left(\frac{b}{a}\right)^2}\,.
           \label{eq;b-6}
\end{equation}
We consider a function $Z(s)$ which is a sum over the $L$ covers from 
$d/2$ to $L_0$. There is a branch point in integrand 
at $z=\pm A_{\rm L}$. Now, we replace the
contour $F_1$ so that it is parallel to the imaginary axis, in order
to deal with the poles of $\tan(\pi z)$. The displacing contour 
$F_2$ is defined as running above the cut associated with
 $z=\pm A_{\rm L}$.
We then obtain an expression for $Z(s)$ by carrying out the 
same calculation as in Appendix \ref{sec:scalar}:
%
\begin{eqnarray}
Z(s)&=&(a)^{2s}\sum^{L_0}_{L=d/2}D_d(L)\nonumber\\
       & &\times \left\{
        \cos(\pi s)\frac{1}{\Gamma(s)}(-1)^{n/2}\frac{1}{2}
        \sum^{n/2-1}_{k=0}f_{nk}\left(A_{\rm L}^2\right)^{k+1-s}
       \Gamma(k+1)\Gamma(s-k-1)
       \right.\nonumber\\
    & &+\cos(\pi s)\int^{\infty}_0 dx\;i\:D_n(ix)
       \left(x^2+A_{\rm L}^2\right)^{-s}
       \frac{2}{{\rm e}^{2\pi x}-1}
      \nonumber\\
    & &\left.+\sin(\pi s)\int^{A_{\rm L}}_0 dx D_n(x)
        \left(A_{\rm L}^2-x^2\right)^{-s}\tan(\pi x)\right\}\,. 
           \label{eq;b-7}
\end{eqnarray}
Here, $D_n(ix)$ is a polynomial with coefficients $f_{Nk}$:
%
\begin{eqnarray}
D_n(ix)&=&i(-1)^{n/2-1}\frac{x}{(n-1)!}
        \left\{x^2+\left(\frac{n}{2}-1\right)^2\right\}\cdots
        \left\{x^2+1\right\}\nonumber\\
       &\equiv&i(-1)^{n/2-1}\sum^{n/2-1}_{k=0}\;f_{nk}\;x^{2k+1}\,.
       \label{eq;b-8}
\end{eqnarray}
Next, we calculate the function $W(s)$.
The branch points in the integrand are on the imaginary axis at 
$z=\pm iB_{\rm L}$. Replacing the contour $F_1$ by $F_3$, we obtain
%
\begin{eqnarray}
W(s)&=&(a)^{2s}\sum^{\infty}_{L=L_0+1}D_d(L)
      \left[\cos(\pi s)\int^{\infty}_{B_{\rm L}} dx\;iD_n(ix)
      \left(x^2+B_{\rm L}^2\right)^{-s}\frac{-2}
      {{\rm e}^{2\pi x}+1}\right.\nonumber\\
     & &+\int^{B_{\rm L}}_{0} dx\;iD_n(ix)
      \left(B_{\rm L}^2-x^2\right)^{-s}
     \frac{-2}{{\rm e}^{2\pi x}+1}\nonumber\\
     & &+\frac{1}{2}(-1)^{n/2}\sum^{n/2-1}_{p=0}
        f_{np}\left(B_{\rm L}^2\right)^{p+1-s}
       \left\{\frac{\Gamma(1-s)\Gamma(s-p-1)}{\Gamma(-p)}\cos(\pi s)
       \right.\nonumber\\
     & &\left.\left.+\frac{\Gamma(1-s)\Gamma(p+1)}{\Gamma(-s+p+2)}\right\}
       \right]\,.
          \label{eq;b-9}
\end{eqnarray}
\begin{figure}[t]
            \epsfxsize = 10 cm
            \epsfysize = 8 cm
            \centerline{\epsfbox{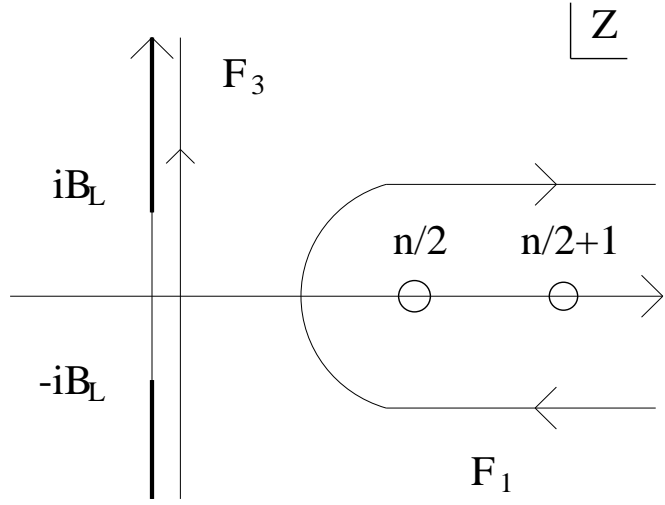}}
\caption{The contour $F_1$ in (\ref{eq;b-9}) replaced by the contour
 $F_3$. The path of $F_3$ runs parallel to the imaginary axis.}
        \label{fig:f2}
\end{figure}
\noindent
The function $W(s)$ contains an infinite sum over $L$. As far as the
first term of Eq.\,(\ref{eq;b-9}) is concerned, 
the infinite sum over $L$ is
convergent, due to the fact that the 
integral decreases exponentially as $L$ goes to
infinity. Contrastingly,
for the first term in (\ref{eq;b-9}), we must regularize 
the summation. First we consider the sum 
%
\begin{equation}
\sum^{\infty}_{L=L_0+1}D_d(L)\left(B_{\rm L}^2\right)^{p+1-s}
             =\left(\frac{a}{b}\right)^{2(p-s+1)}
           \sum^{\infty}_{L=L_0+1}D_d(L)\left(L^2-C^2\right)^{-\beta}\,,
           \label{eq;b-10}
\end{equation}
with
%
\begin{equation}
C^2=\frac{n(n-1)}{4}\left(\frac{b}{a}\right)^2\,,
           \label{eq;b-11}
\end{equation}
where $\beta=s-p-1$. Then, using Eq.\,(\ref{eq;b-9}), the function 
$W(s)$ is given by
%
\begin{eqnarray}
W(s)&=&(a)^{2s}\sum^{\infty}_{L=L_0+1}D_d(L)
      \left[\cos(\pi s)\int^{\infty}_{B_{\rm L}} dx\;iD_n(ix)
      \left(x^2+B_{\rm L}^2\right)^{-s}
       \frac{-2}{{\rm e}^{2\pi x}+1}\right.
      \nonumber\\
    & &+\left.\int^{B_{\rm L}}_{0} dx\;iD_n(ix)
      \left(B_{\rm L}^2-x^2\right)^{-s}\frac{-2}
      {{\rm e}^{2\pi x}+1}\right]\nonumber\\
    & &+(a)^{2s}\frac{1}{2}(-1)^{n/2}\sum^{n/2-1}_{p=0}
        C_{np}\left(\frac{a}{b}\right)^{2(p+1-s)}
       \left\{\frac{\Gamma(1-s)\Gamma(s-p-1)}{\Gamma(-p)}\cos(\pi s)
        \right.\nonumber\\
    & &\left.+\frac{\Gamma(1-s)\Gamma(p+1)}{\Gamma(-s+p+2)}\right\}
     \sum^{\infty}_{L=L_0+1}D_d(L)\left(L^2-C^2\right)^{p+1-s}\,.         
           \label{eq;b-12}
\end{eqnarray}
For the last term in Eq.\,(\ref{eq;b-12}), 
we replace the sum by the following integration: 
%
\begin{equation}
S\equiv\sum^{\infty}_{L=L_0+1}D_d(L)\left(L^2-C^2\right)^{-\beta}
   =\frac{-1}{2i}\int_{F_4}dz D_d(z)\left(z^2-C^2\right)^{-\beta}
    \cot(\pi z)\,.
           \label{eq;b-13}
\end{equation}
The contour is the same as that for Eq.\,(\ref{eq;a-15})
 in Appendix \ref{sec:scalar}.
By replacing the contour $F_4$ by $F_5$, we obtain 
%
\begin{eqnarray}
S&=&-\sin(\pi \beta)\left\{\int^{\infty}_{0}dx\;
    D_d(ix)\left(x^2+C^2\right)^{-\beta}
    \left(\frac{2}{{\rm e}^{2\pi x}-1}\right)\right.\nonumber\\
   & + &P\int^{C}_{0}dx\;D_d(x)\left(C^2-x^2\right)^{-\beta}
    \cot(\pi x)\nonumber\\
   & + & \left.(-1)^{d/2}\frac{1}{2}
    \sum^{d/2-1}_{q=0}\;f_{dq}\;\left(C^2\right)^{q-\beta+\frac{3}{2}}
    \frac{\Gamma\left(q+\frac{3}{2}\right)
    \Gamma\left(\beta-q-\frac{3}{2}\right)}{\Gamma\left(\beta\right)}
   \right\}\,,
           \label{eq;b-14}
\end{eqnarray}
where $\beta=s-r-1$, the function 
$D_d(ix)$ is defined as 
%
\begin{eqnarray}
D_d(ix)&=&(-1)^{(d-1)/2}\frac{2x^2}{d!}
        \left\{x^2+\left(\frac{d}{2}-1\right)^2\right\}\cdots
        \left\{x^2+\left(\frac{1}{2}\right)^2\right\}\nonumber\\
       &\equiv&(-1)^{(d-1)/2}\sum^{(d-1)/2-1}_{q=0}\;f_{dq}\;x^{2q}\,,
             \label{eq;b-15}
\end{eqnarray}
$P$ stands for the principal value integral, and 
the coefficients $f_{dq}$ in $D_{d}$ are defined in Eq.\,(\ref{eq;b-15}).
\begin{figure}[t]
            \epsfxsize = 9 cm
            \epsfysize = 7.5 cm
            \centerline{\epsfbox{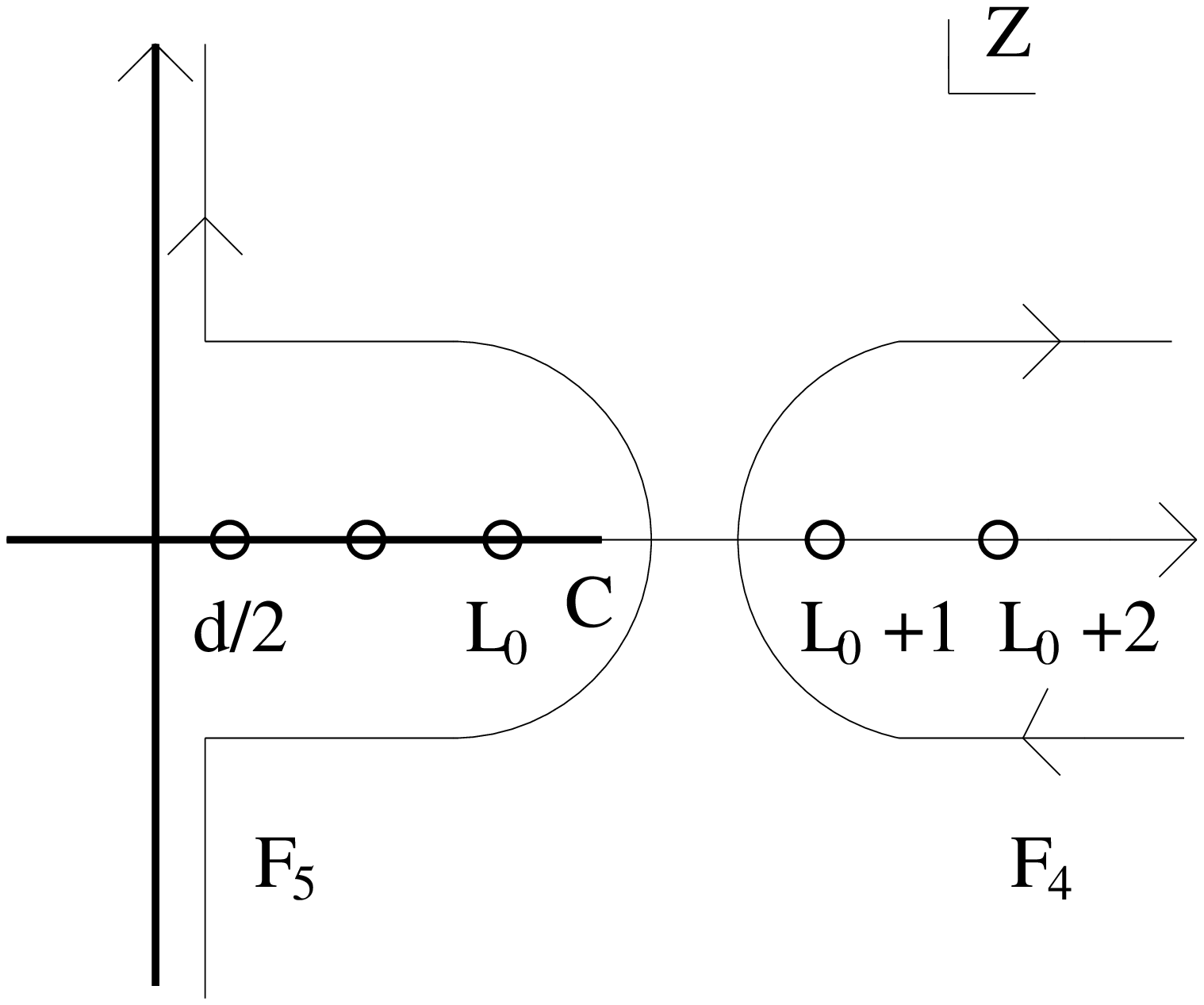}}
\caption{The contour $F_4$ in (\ref{eq;b-15}) replaced by the contour
 $F_5$. The pole of $\cos(\pi z)$ exists at $z=L_0$, which is a smaller
 value than that at the cut of the branch point.}
        \label{fig:f3}
\end{figure}
\noindent
Substituting (\ref{eq;b-14}) into the last term of 
Eq.\,(\ref{eq;b-12}), we obtain an expression for $W(s)$.
Note that the value of $S$ at $s=0$ is zero and thus does
 not contribute to 
 $\zeta(0)$. Then, $W(s)$, is given by
%
\begin{eqnarray}
W(s)&=&(a)^{2s}\sum^{\infty}_{L=L_0+1}D_d(L)
      \left[\cos(\pi s)\int^{\infty}_{B_{\rm L}} dx\;iD_n(ix)
      \left(x^2+B_{\rm L}^2\right)^{-s}
       \frac{-2}{{\rm e}^{2\pi x}+1}\right.
      \nonumber\\
     & &+\left.\int^{B_{\rm L}}_{0} dx\;iD_n(ix)
     \left(B_{\rm L}^2-x^2\right)^{-s}\frac{-2}
     {{\rm e}^{2\pi x}+1}\right]\nonumber\\
     & &+(a)^{2s}\frac{1}{2}(-1)^{n/2}\sum^{n/2-1}_{p=0}
        f_{np}\left(\frac{a}{b}\right)^{2(p+1-s)}
       \left\{\frac{\Gamma(1-s)\Gamma(s-p-1)}{\Gamma(-p)}\cos(\pi s)
         \right.\nonumber\\
     & &\left.+\frac{\Gamma(1-s)\Gamma(p+1)}{\Gamma(-s+p+2)}\right\}
       \times S\,.\nonumber\\
           \label{eq;b-16}         
\end{eqnarray}
Finally, we find that the generalized $\zeta$ function is expressed as 
%
\begin{eqnarray}
\zeta(s)&=&(a)^{2s}\sum^{L_0}_{L=d/2}D_d(L)\nonumber\\
        & &\times\left\{
        \cos(\pi s)\frac{1}{\Gamma(s)}(-1)^{n/2}\frac{1}{2}
        \sum^{n/2-1}_{k=0}f_{nk}\left(A_{\rm L}^2\right)^{k+1-s}
       \Gamma(k+1)\Gamma(s-k-1)
       \right.\nonumber\\
    & &+\cos(\pi s)\int^{\infty}_0 dx\;i\:D_n(ix)
       \left(x^2+A_{\rm L}^2\right)^{-s}
       \frac{2}{{\rm e}^{2\pi x}-1}
      \nonumber\\
    & &+\left.\sin(\pi s)\int^{A_{\rm L}}_0 dx D_n(x)
        \left(A_{\rm L}^2-x^2\right)^{-s}\tan(\pi x)\right\}\nonumber\\
    & &+(a)^{2s}\sum^{\infty}_{L=L_0+1}D_d(L)
      \left[\cos(\pi s)\int^{\infty}_{B_{\rm L}} dx\;iD_n(ix)
      \left(x^2+B_{\rm L}^2\right)^{-s}
      \frac{-2}{{\rm e}^{2\pi x}+1}\right.
      \nonumber\\      
    & &+\left.\int^{B_{\rm L}}_{0} dx\;iD_n(ix)
      \left(B_{\rm L}^2-x^2\right)\frac{-2}
       {{\rm e}^{2\pi x}+1}\right]\nonumber\\
     & &+(a)^{2s}\frac{1}{2}(-1)^{n/2}\sum^{n/2-1}_{p=0}
        f_{np}\left(\frac{a}{b_0}\right)^{p+1-2s}
       \left\{\frac{\Gamma(1-s)\Gamma(s-p-1)}{\Gamma(-p)}\cos(\pi s)
       \right.\nonumber\\
      & & \left.+\frac{\Gamma(1-s)\Gamma(p+1)}{\Gamma(-s+p+2)}\right\}
       \times S\,.\nonumber\\ 
           \label{eq;b-17}
\end{eqnarray}

%% file: appenC.tex
\section{Zeta Function Regularization for a Vector Field}
         \label{sec:vector}
\begin{figure}[t]
            \epsfxsize = 10 cm
            \epsfysize = 8 cm
            \centerline{\epsfbox{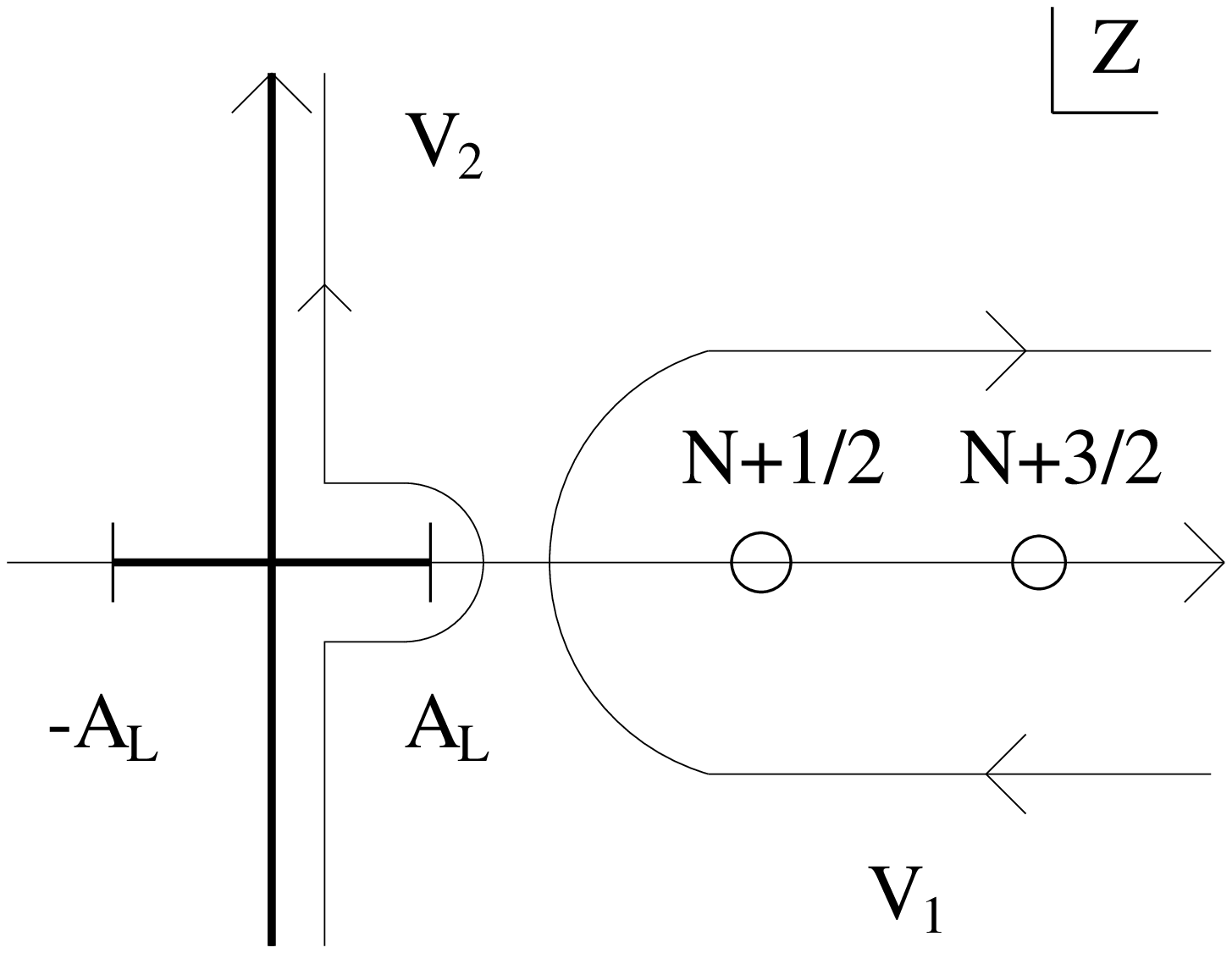}}
\caption{The contour $V_1$ in Eq.\,(\ref{eq;c-4}) is replaced by the
 contour $V_2$. Note that the contour $V_2$ avoids the branch points at
 $z=\pm A_{\rm L}$.}
        \label{fig:v1}
\end{figure}

In this appendix, we present a method to regularize the zeta function
for a vector field in an odd total number of dimensions. 
We consider the case of the field $A_{{\rm (T)}\;\mu}$.  The quantity
$A_{\rm (T)}$ can be calculated in a similar way, and   
this method can be applied to the case of an 
even number of dimensions also. 
We define a generalized zeta function for the
vector fields $A_{{\rm (T)}\:\mu}$ in $S^n\times S^d$ as 
%
\begin{eqnarray}
\zeta(s)&=&\sum^{\infty}_{l'=1}\sum^{\infty}_{l=0}
         \frac{(l'+n-3)!\;(l+d-2)!}{(n-2)!\;(d-1)!}
         \frac{l'\;(l'+n-1)\;(2l'+n-1)\;(2l+d-1)}{(l'+1)!\;l!}
            \nonumber\\
        & & \hspace{0.5cm}
            \times\left\{\frac{l'(l'+d-1)-1-
             \left(\frac{n-1}{2}\right)}{a^2}
            +\frac{l(l+d-1)}{b_0^2}
             \left(\frac{b_0}{b}\right)^{d+2}\right\}^{-s}\,,
           \label{eq;c-1}
\end{eqnarray}
where $n$ is an even number and $d$ is an odd number.
Using $N=n/2$ and $D=(d-1)/2$ instead of $n$ and $d$ and running variables
 $L'=l'+N$ and $L=l+D$, we rewrite (\ref{eq;c-1}) as  
%
\begin{equation}
\zeta(s)
         =\sum^{\infty}_{L'=N}D_{n}\left(L'-\frac{1}{2}\right)
         \sum^{\infty}_{L=D}D_d(L)
         \left\{\frac{\Lambda_n\left(L'-\frac{1}{2}\right)}{a^2}
         +\frac{\Lambda_d(L)}{b_0^2}
         \left(\frac{b_0}{b}\right)^{d+2}\right\}^{-s}\,, 
           \label{eq;c-2}
\end{equation}
where the eigenvalues and degeneracies are now given by
%
\begin{eqnarray}
D_{n}\left(L'-\frac{1}{2}\right)&=&
             \frac{2L'-1}{(2N-1)!}\left\{\left(L'-\frac{1}{2}\right)^2
             -\left(N-\frac{1}{2}\right)^2\right\}\nonumber\\  
             &\times&
             \left\{\left(L'-\frac{1}{2}\right)^2
             -\left(N-\frac{5}{2}\right)^2\right\}
             \cdots
             \left\{\left(L'-\frac{1}{2}\right)^2
             -\left(\frac{1}{2}\right)^2\right\}\,,\nonumber\\
D_{d}(L)&=&\frac{2L^2}{(2D)!}\left\{L^2-(D-1)^2\right\}
            \cdots\left\{L^2-1\right\},\nonumber\\
\Lambda_{n}
             \left(L'-\frac{1}{2}\right)&=&\left(L'-\frac{1}{2}\right)^2
             -\left(N-\frac{1}{2}\right)^2-N-\frac{1}{2},\nonumber\\
\Lambda_d(L)&=&L^2-D^2\,.
           \label{eq;c-3}
\end{eqnarray}
Next, replacing the infinite mode sum over $L'$ by complex integration,  
the generalized zeta function is  
%
\begin{eqnarray}
\zeta(s)&=&
         (a)^{2s}\sum^{\infty}_{L=D}D_d(L)\frac{-1}{2i}\int_{V_1}dz
         \tan(\pi z)D_n(z)
            \nonumber\\
         &\times&\left\{z^2+\left(\frac{a}{b_0}\right)^2
         \left(\frac{b_0}{b}\right)^{d+2}\Lambda_d(L)
         -\left(N-\frac{1}{2}\right)^2
         -N-\frac{1}{2}\right\}^{-s}\,,
           \label{eq;c-4}
\end{eqnarray}
where the contour $V_1$ in the complex plane is showed in 
Fig.\,\ref{fig:v1}.
\noindent
In order to avoid the singularity of $z$ in the integrand, it is
convenient to divide the sum over $L$ into two parts as
%
\begin{eqnarray}
\zeta(s)=Z(s)+W(s)\,,
           \label{eq;c-5}
\end{eqnarray}
where
%
\begin{eqnarray}
Z(s)&=&(a)^{2s}\sum^{L_0}_{L=D}D_d(L)\frac{-i}{2}\int_{V_1}dz
         \tan(\pi z)D_n(z)\left(z^2-A_{\rm L}^2\right)^{-s}\,,\nonumber\\
W(s)&=&(a)^{2s}\sum^{\infty}_{L=L_0+1}D_d(L)\frac{-i}{2}\int_{V_1}dz
         \tan(\pi z)D_n(z)\left(z^2+B_{\rm L}^2\right)^{-s}\,,
           \label{eq;c-6}
\end{eqnarray}
with
%
\begin{eqnarray}
A_{\rm L}^2&=&\left(N-\frac{1}{2}\right)^2+N+\frac{1}{2}
      -\left(\frac{a}{b_0}\right)^2\left(\frac{b_0}{b}\right)^{d+2}
      \Lambda_d(L)\,,\nonumber\\
B_{\rm L}^2&=&-A_{\rm (T)\;L}^2\,.\nonumber\\
           \label{eq;c-7}
\end{eqnarray}
Here, $A_{\rm L}^2$ and $B_{\rm L}^2$
 are positive for $D\le L<L_0$ and $L_0\le L$,
 respectively. $L_0$ is defined as the largest integer smaller than
 or equal to 
%
\begin{equation}
\left[D^2+\left(\frac{b}{a}\right)^2\left(\frac{b}{b_0}\right)^{d+2}
\left\{\left(N-\frac{1}{2}\right)^2+N+\frac{1}{2}\right\}\right]^{1/2}\,.
           \label{eq;c-8}
\end{equation}
We consider the function $Z(s)$ which is sum over the $L$ covers from 
$D$ to $L_0$. There are branch points at $z=\pm A_{\rm L}$.
 Now, we replace the
contour $V_1$ so that it is along to the imaginary axis in order
to deal with the poles of $\cot(\pi z)$. Since the displacing contour 
$V_2$ is defined as running above the cut associated with 
$z=\pm A_{\rm L}$,
the result of $Z(s)$ is given by  
%
\begin{eqnarray}
Z(s)&=&(a)^{2s}\sum^{L_0}_{L=D}D_d(L)\left\{\frac{1}{2}
      \left({\rm e}^{-i\pi s}
    +{\rm e}^{i\pi s}\right)\int^{\infty}_0dx\;i\:D_n(ix)
      \left(x^2+A_{\rm L}^2\right)^{-s}\tanh(\pi x)\right.\nonumber\\
    & &-\frac{1}{2}{\rm e}^{i\pi s}\int^{A_L}_0
      dx\:\tan\pi(x-i\epsilon)D_n(x)\left(A_{\rm L}^2-x^2\right)^{-s}
      \nonumber\\
    & &\left.-\frac{1}{2}{\rm e}^{-i\pi s}\int^{A_L}_0
      dx\:\tan\pi(x+i\epsilon)D_n(x)\left(A_{\rm L}^2-x^2\right)^{-s}
      \right\}\,,
           \label{eq;c-9}
\end{eqnarray}
where $D_n(ix)$ is a polynomial with coefficients $j_{Nk}$:
%
\begin{eqnarray}
D_n(ix)&=&i(-1)^{N-1}\frac{2x}{(2N-1)!}
        \left\{x^2+\left(N-\frac{1}{2}\right)^2\right\}\nonumber\\
       & &\times \left\{x^2+\left(N-\frac{5}{2}\right)^2\right\}
       \cdots
        \left\{x^2+\left(\frac{1}{2}\right)^2\right\}\nonumber\\
       &\equiv&i(-1)^{N-1}\sum^{N-1}_{k=1}\;j_{Nk}\;x^{2k+1}\,.
           \label{eq;c-10}
\end{eqnarray}
The first term in Eq.\,(\ref{eq;c-9}) comes from the integral 
along the imaginary
axis. The second and third terms in Eq.\,(\ref{eq;c-9}) 
are the contributions from the
contour along the cut on the real axis.
Since we use the relation 
%
\begin{equation}
\tanh(\pi x)=1-\frac{2}{{\rm e}^{2\pi x}+1}\,,
           \label{eq;c-11}
\end{equation}
and we substitute this into the first term, the function $Z(s)$ is   
%
\begin{eqnarray}
Z(s)&=&-(a)^{2s}\sum^{L_0}_{L=D}D_d(L)\nonumber\\
    & &\times\left\{\cos(\pi s)
      \frac{1}{\Gamma(s)}(-1)^N\frac{1}{2}\sum^{N-1}_{k=1}j_{Nk}
      \left(A_{\rm L}^2\right)^{k+1-s}\Gamma(k+1)\Gamma(s-k-1)\right.
      \nonumber\\
    & &+\cos(\pi s)\int^{\infty}_0 dx\;i\:D_n(ix)
       \left(x^2+A_{\rm L}^2\right)^{-s}
       \frac{2}{{\rm e}^{2\pi x}-1}
      \nonumber\\
    & &\left.+\sin(\pi s)\int^{A_{\rm L}}_0 dx D_n(x)
        \left(A_{\rm L}^2-x^2\right)^{-s}\tan(\pi x)\right\}\,. 
           \label{eq;c-12}
\end{eqnarray}
We note that first term in Eq.\,(\ref{eq;c-12})
 does not contribute to $\zeta(s)$.

We next consider the function $W(s)$. 
The branch points in integrand are on
the imaginary axis at $z=\pm B_{\rm L}$.
Replacing the contour $V_1$ by $V_3$, we obtain
%
\begin{eqnarray}
W(s)&=&-\sum^{\infty}_{L=L_0+1}D_d(L)(a)^{2s}\cos(\pi s)\nonumber\\
    & &\times\left\{
      \frac{1}{\Gamma(s)}(-1)^N\frac{1}{2}\sum^{N-1}_{p=0}j_{Np}
      \left(B_{\rm L}^2\right)^{p+1-s}\Gamma(s-p-1)\Gamma(p+1) \right.
           \nonumber\\
    & &\left.+\int^{\infty}_{0} dx\:i\:D_n(ix)
       \left(x^2-B_{\rm L}^2\right)^{-s}
       \frac{2}{{\rm e}^{2\pi x}+1}\right\}\,.
           \label{eq;c-13}
\end{eqnarray}
\begin{figure}[t]
            \epsfxsize = 10 cm
            \epsfysize = 8 cm
            \centerline{\epsfbox{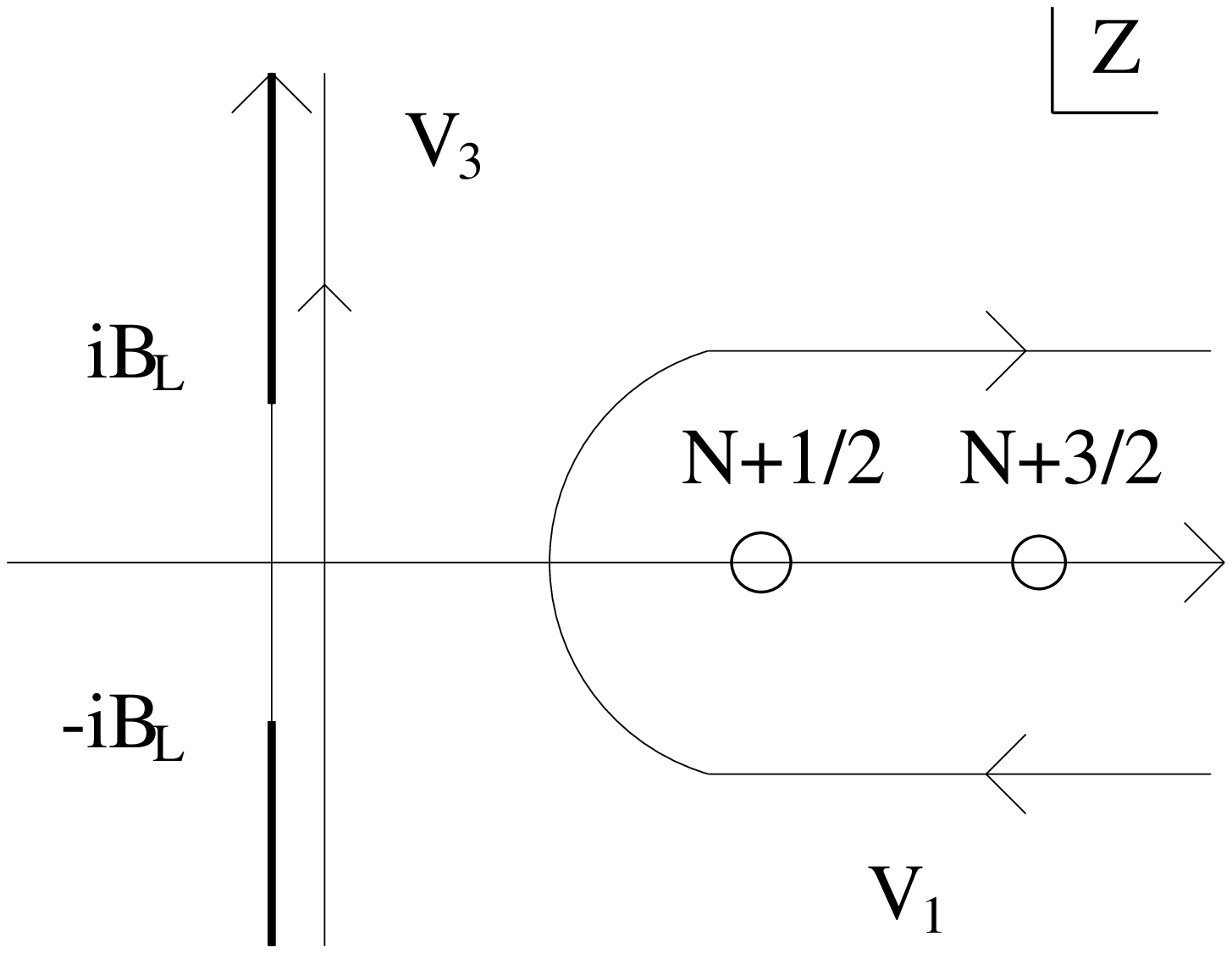}}
\caption{The contour $V_1$ in (\ref{eq;c-13}) replaced by the contour
 $V_3$. The path of $V_3$ runs parallel to the imaginary axis.}
        \label{fig:v2}
\end{figure}
\noindent
The function $W(s)$ is contains an infinite sum over $L$. As far as the
first term of Eq.\,(\ref{eq;c-13})
 is concerned, the infinite sum over $L$ is
convergent, due to the fact that 
the integral decreases exponentially as $L$ goes to
infinity. For the second term in Eq.\,(\ref{eq;c-13}), 
however, we must regularize
the summation  
%
\begin{equation}
\sum^{\infty}_{L=L_0+1}D_d(L)\left(B_{\rm L}^2\right)^{p+1-s}
             =\left\{\left(\frac{a}{b_0}\right)^{2}
              \left(\frac{b_0}{b}\right)^{d+2}\right\}^{p-s+1}
           \sum^{\infty}_{L=L_0+1}D_d(L)\left(L^2-C^2\right)^{-\beta}\,,
           \label{eq;c-14}
\end{equation}
with 
%
\begin{equation}
C^2=D^2+\left(\frac{b_0}{a}\right)^2\left(\frac{b}{b_0}\right)^{d+2}
    \left\{\left(N-\frac{1}{2}\right)^2+N+\frac{1}{2}\right\}\,,
           \label{eq;c-15}
\end{equation}
where $\beta=s-p-1$. 
We again replace the infinite sum with a complex integral and thereby
obtain, the integral is  
%
\begin{equation}
S=\sum^{\infty}_{L=L_0+1}D_d(L)\left(L^2-C^2\right)^{-\beta}
 =\frac{-1}{2i}\int_{C_3}dz D_d(z)\left(z^2-C^2\right)^{-\beta}
    \cot(\pi z)\,,
           \label{eq;c-16}
\end{equation}
where the sum is taken over half integers,
 and the contour $V_4$ is depicted in Fig.\,\ref{fig:v2}.
 The function $D_d(ix)$ is defined as 
%
\begin{eqnarray}
D_d(ix)&=&(-1)^{D}\frac{2x^2}{(2D)!}
        \left\{x^2+\left(D-1\right)^2\right\}\cdots
        \left\{x^2+1^2\right\}\nonumber\\
       &\equiv&(-1)^{D}\sum^{D-1}_{q=0}\;j_{Dq}\;x^{2q+2}\,.
           \label{eq;c-17}
\end{eqnarray}
Replacing the contour $V_4$ by $V_5$, we obtain 
%
\begin{eqnarray}
S&=&-\cos(\pi \beta)\sum^{L_0}_{L=D}D_d(L)\left(C^2-L^2\right)^{-\beta}
    \nonumber\\
 & &+\sin(\pi \beta)\left\{\int^{\infty}_{0}dx\;i\:
    D_n(ix)\left(x^2+C^2\right)^{-\beta}
    \left(\frac{1}{{\rm e}^{2\pi x}-1}\right)\right.\nonumber\\
 & &+\int^{C}_{0}dx\;D_n(x)\left(C^2-x^2\right)^{-\beta}
    \cot(\pi x)\nonumber\\
 & &\left.+(-1)^{D}\frac{1}{2}
    \sum^{D-1}_{q=0}\;j_{Dq}\;\left(C^2\right)^{q-\beta+\frac{3}{2}}
    \frac{\Gamma\left(q+\frac{3}{2}\right)
    \Gamma\left(\beta-q-\frac{3}{2}\right)}{\Gamma\left(\beta\right)}
   \right\}\,,
           \label{eq;c-18}
\end{eqnarray}
where $P$ stands for the principal value integral, and 
the coefficients $j_{Dq}$ in $D_{d}$ are defined in Eq.\,(\ref{eq;c-17}).
\begin{figure}[t]
            \epsfxsize = 9 cm
            \epsfysize = 7.5 cm
            \centerline{\epsfbox{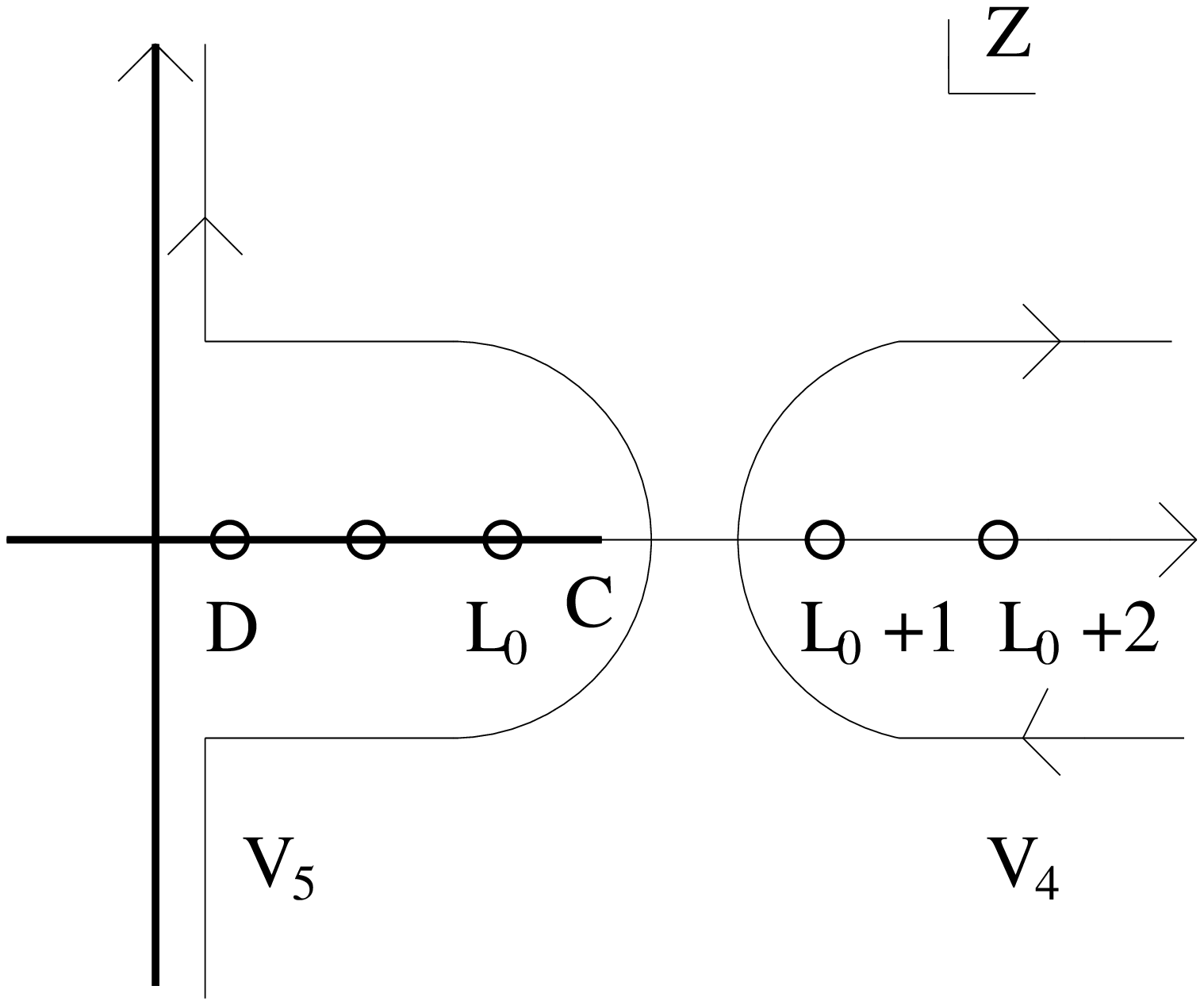}}
\caption{The contour $V_4$ in (\ref{eq;c-18}) replaced by the contour
 $V_5$. The pole of $\cos(\pi z)$ exists at $z=L_0$ which is smaller
 value  than that at the cut of the branch point.}
        \label{fig:v3}
\end{figure}
\noindent
Substituting Eq.\,(\ref{eq;c-18}) into the second term of
 Eq.\,(\ref{eq;c-13}), we obtain an expression for
 $W(s)$. 
Finally, the generalized $\zeta$ function is found to be  
%
\begin{eqnarray}
\zeta(s)&=&(a)^{2s}\sum^{L_0}_{L=D}D_d(L)\nonumber\\
    & &\times\left[\cos(\pi s)\left\{
      \frac{1}{\Gamma(s)}(-1)^N\frac{1}{2}\sum^{\infty}_{k=0}j_{Nk}
      \left(A_{\rm L}^2\right)^{k-s+1}\Gamma(k+1)\;\Gamma(s-k-1) 
        \right.\right.\nonumber\\
    & &\left.
       -\int^{\infty}_0 dx\;i\:D_n(ix)\left(x^2+A_{\rm L}^2\right)^{-s}
       \frac{2}{{\rm e}^{2\pi x}+1}\right\}\nonumber\\
    & &-\left.\sin(\pi s)
       \int^{A_{\rm L}}_0 dx D(x)
      \left(A_{\rm L}^2-x^2\right)^{-s}\tan(\pi x)
      \right]\nonumber\\
    & &-(a)^{2s}\sum^{\infty}_{L=L_0+1}D_d(L)\;\cos(\pi s)\left\{
       \int^{\infty}_{B_L} dx\;i\:D_n(ix)\left(x^2-B_L^2\right)^{-s} 
       \frac{2}{{\rm e}^{2\pi x}+1}\right\}\nonumber\\
    & &+\cos(\pi s)(a)^{2s}\Gamma(-s+1)(-1)^N\frac{1}{2}
       \sum^{N-1}_{p=0}j_{Np}\left\{\left(\frac{a}{b_0}\right)^2
       \left(\frac{b_0}{b}\right)^{d+2}\right\}^{p-s+1}
      \nonumber\\
    & &\times\frac{\Gamma(s-p-1)}{\Gamma(-p)}\; S\,.   
           \label{eq;c-19}
\end{eqnarray}

%% file: appenD.tex
\section{Zeta Function Regularization of One-Loop Effective Potential in
 $AdS_n\times S^d$}
         \label{sec:zads}
In this appendix, we present a method to calculate the zeta
function regularization for a scalar field in the product spacetime 
 $AdS_n\times S^d$. 
The Euclidean section for the $AdS_n$ spacetime is the
$n$-dimensional hyperbolic space $H^n$. The calculation of the 
zeta function
for $AdS_n$ is discussed in Ref.\,40).
We extend the
calculation technique given them to
 the zeta function for $AdS_n\times S^d$.    
On a compact Euclidean section, the zeta function is given by 
%
\begin{equation}
\zeta_{\phi}(s)=\sum^{\infty}_{l=0}D_l\;\Lambda_l^{-s}\,, 
\end{equation}
where $\Lambda_l$ is the discrete eigenvalue of the Laplace-Beltrami
operator and $D_l$ is the degeneracy of the eigenvalue.
The calculation of the zeta function on $S^d$ is performed using the 
well-known spectrum of the Laplace-Beltrami operator on $S^d$.
The zeta function for a noncompact manifold, however, is not the 
same as that in the compact case. In the homogeneous $n$-dimensional
hyperbolic space $H^n$, the zeta function takes the form\cite{cam1}
%
\begin{equation}
\zeta_{\phi}(s)=\int^{\infty}_0 d\lambda\;\mu(\lambda)\; 
               \Lambda(\lambda)^{-s}\,, 
\end{equation}
where $\Lambda(\lambda)$ is the eigenvalue of the Laplace-Beltrami
operator on $H^n$, $\lambda$ is the real parameter that labels
the continuous spectrum, and $\mu(\lambda)$ is the spectrum, function (or
Plancherel measure) on $H^n$ corresponding to the discrete degeneracy on
$S^d$. 
The spectrum function for the scalar and spinor fields
on $H^n$ is calculated in Refs.\,41) and 42).
Also, the
spectrum function for the transverse-traceless tensor field 
on $H^n$ is given in Ref.\,43).

\begin{figure}[t]
            \epsfxsize = 10 cm
            \epsfysize = 8 cm
            \centerline{\epsfbox{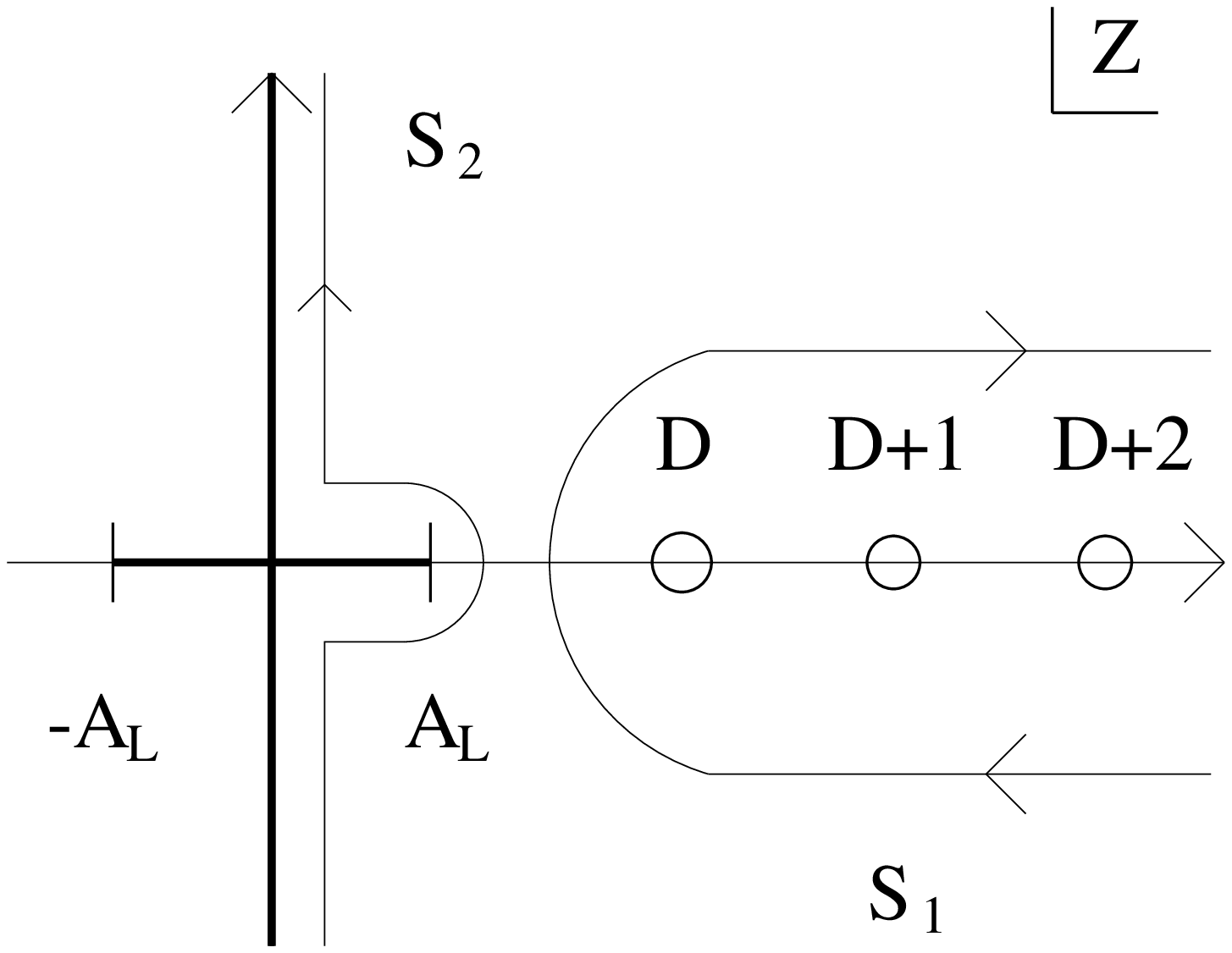}}
\caption{The contour $S_1$ in Eq.\,(\ref{eq;d-6}) is replaced by the
 contour $S_2$. Note that the contour $S_2$ avoids the branch points at
 $z=\pm A_{\rm L}$.}
        \label{fig:d1}
\end{figure}

\begin{figure}[t]
            \epsfxsize = 10 cm
            \epsfysize = 8 cm
            \centerline{\epsfbox{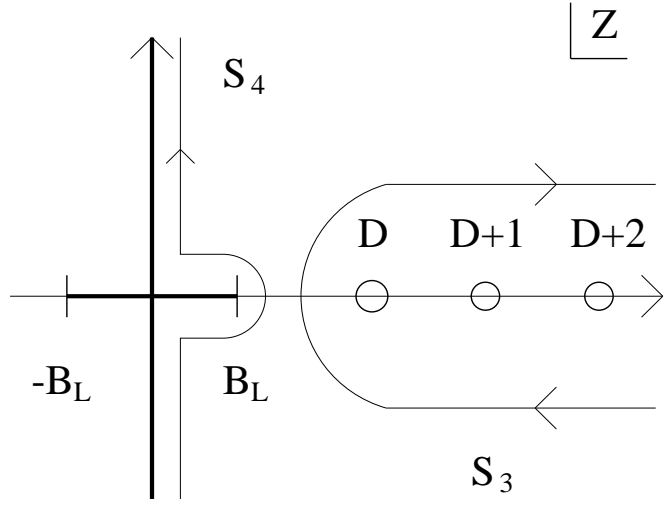}}
\caption{The contour $S_3$ in Eq.\,(\ref{eq;d-6}) is replaced by the
 contour $S_4$. Note that the contour $S_4$ avoids the branch points at
 $z=\pm B_{\rm L}$.}
        \label{fig:d2}
\end{figure}

 Here, we consider the zeta function regularization of the one-loop 
effective potential for scalar field in $AdS_n\times S^d$.
This regularization scheme can be extended for other fields.
Define the generalized zeta function in $H^n\times S^d$
for the scalar field as  
%
\begin{eqnarray}
\zeta_{\phi}(s)&=&\sum^{\infty}_{l=0}
         \frac{(l+d-2)!}{(d-1)!}
         \frac{(2l+d-1)}{l!}
         \int^{\infty}_0 d\lambda\;\mu(\lambda)
            \nonumber\\
        & & \hspace{0.5cm}
            \times\left\{\frac{\lambda^2+\{(n-1)/2\}^2}{a^2}
            +\frac{l(l+d-1)}{b_0^2}
                {\rm e}^{-[2d/(n-2)+2]\kappa_\sigma \sigma}\right\}^{-s}\,.
       \label{eq;d-1}
\end{eqnarray}
For even dimensions,
the Plancherel measure $\mu(\lambda)$ is given by\cite{cam1}
%
\begin{equation}
\mu(\lambda)=\frac{\pi\lambda \tanh(\pi \lambda)}
    {2^{2(n-2)}\left\{\Gamma(n/2)\right\}^2}
             \prod^{(n-3)/2}_{j=1/2}\left(\lambda^2+j^2\right)\,.
       \label{eq;d-2}
\end{equation}
Then,
 using $D=(d-1)/2$ and 
$N=(n-1)/2$ instead of $d$ and $n$ and the running variables
$L=l+D$, we rewrite (\ref{eq;d-1}) as  
%
\begin{equation}
\zeta_{\phi}(s)=\sum^{\infty}_{L=D}D_{\phi}(L)\int^{\infty}_0 
           d\lambda\;\mu(\lambda)
         \left(\frac{\lambda^2+N^2}{a^2}
         +M_{\phi}^2\right)^{-s}\,, 
         \label{eq;d-3}
\end{equation}
where the $D_{\phi}(L)$ and  $\Lambda_{\phi}(L)$ are given by
%
\begin{eqnarray}
D_{\phi}(L)&=&\frac{2L^2}{(2D)!}\left\{L^2-(D-1)^2\right\}
            \cdots\left\{L^2-1\right\},\nonumber\\
\Lambda_{\phi}(L)&=&L^2-D^2,\nonumber\\
M_{\phi}^2&=&\frac{\Lambda_{\phi}(L)}{b_0^2}
                 {\rm e}^{-[2d/(n-2)+2]\kappa_\sigma \sigma}\,.
        \label{eq;d-4}
\end{eqnarray}
Next, integrating Eq.\,(\ref{eq;d-3}) over
 $\lambda$, we find the expression   
%
\begin{eqnarray}
\zeta_{\phi}(s)&=&\frac{\pi}{2^{2n-1}\left\{\Gamma(2N+1)\right\}^2}
        \frac{a^{2s}}{\Gamma(s)}\sum^{\infty}_{L=D}D_{\phi}(L)
        \nonumber\\ 
        & &\times
        \left[\frac{\left(N^2+a^2 M^2\right)^{-s+1}}{2(s-1)(s-2)}
        \left\{N^2+a^2 M^2+j^2(s-2)\right\}\right.\nonumber\\
       & &\left.-\int^{\infty}_0 d\lambda~
        \prod^{(n-3)/2}_{j=1/2}\left(\lambda^2+j^2\right)
        \frac{2\lambda}
        {({\rm e}^{2\pi\lambda}+1)\left(\lambda^2+N^2+a^2 M^2\right)}
        \right]\,,
              \label{eq;d-5}
\end{eqnarray}
where we have used the identity
%
\begin{eqnarray}
\tanh(\pi \lambda)=1-\frac{2}{{\rm e}^{2\pi\lambda}+1}\,.
\end{eqnarray}
To regularize the mode sum in Eq.\,(\ref{eq;d-5}), we replace the infinite
sum over $L$ by complex integration.
The generalized zeta function is then
%
\begin{eqnarray}
\zeta_{\phi}(s)&=&\frac{\pi}
       {2^{2n-1}\left\{\Gamma(2N+1)\right\}^2\Gamma(s)}
        \left[\frac{i}{4(s-1)(s-2)}
       \left\{\left(\frac{a}{b_0}\right)^2
        {\rm e}^{-[2d/(n-2)+2]\kappa_\sigma \sigma}
      \right\}^{-s+2}\right.\nonumber\\ 
       & &\times\int_{S_1}dz\;D_{\phi}(z)\:\cot(\pi z)
       \left(z^2-A_{\rm L}^2\right)^{-s+2}\nonumber\\
       & & + \frac{i}{2(s-1)}\left\{\left(\frac{a}{b_0}\right)^2
        {\rm e}^{-[2d/(n-2)+2]\kappa_\sigma \sigma}
       \right\}^{-s+1}\nonumber\\\
       & &\times\int_{S_1}dz\;D_{\phi}(z)\:\cot(\pi z)
       \left(z^2-A_{\rm L}^2\right)^{-s+1}\nonumber\\
        & &\left.-i\int^{\infty}_0 d\lambda 
       \prod^{(n-3)/2}_{j=1/2}\left(\lambda^2+j^2\right)\frac{1}
       {{\rm e}^{2\pi\lambda}+1}\int_{S_3} dz\:\cot(\pi z)
       \left(z^2-B_{\rm L}^2\right)^{-s}\right]\,,
        \label{eq;d-6}
\end{eqnarray}
where  $A^2_{\rm L}$ and $B^2_{\rm L}$ are given by
%
\begin{eqnarray}
A_{\rm L}^2&=&D^2
      -N^2\left(\frac{b_0}{a}\right)^2
              {\rm e}^{[2d/(n-2)+2]\kappa_\sigma \sigma}\,,\nonumber\\
B_{\rm L}^2&=&D^2
      -\left(N^2+j^2\right)\left(\frac{b_0}{a}\right)^2
              {\rm e}^{[2d/(n-2)+2]\kappa_\sigma \sigma}\,.
       \label{eq;d-7}
\end{eqnarray}
and the contours $S_1$ and $S_3$ in the complex plane are showed 
in Figs.\,\ref{fig:d1} and \ref{fig:d2}.
Note that there are two branch points, $z=\pm A_{\rm L}$,
 in the integration.
We introduce the functions $\tilde{\zeta}_k(s)$ and  
$\hat{\zeta}_k(s)$ as 
%
\begin{eqnarray}
\tilde{\zeta}_k(s) &=&
        \frac{\pi a^{2s}}{2^{2n-1}\left\{\Gamma(2N+1)\right\}^2\Gamma(s)}
                \left[\frac{i}{4(s-1)(s-2)}
       \left\{\left(\frac{a}{b_0}\right)^2
        {\rm e}^{-[2d/(n-2)+2]\kappa_\sigma \sigma}
       \right\}^{-s+k}
       \right.\nonumber\\ 
        &\times&\left.\int_{S_1}dz\;D_{\phi}(z)\:\cot(\pi z)
       \left(z^2-A_{\rm L}^2\right)^{-s+k}\right]\,,
        \nonumber\\
\hat{\zeta}_k(s) &=&
        \frac{\pi a^{2s}}{2^{2n-1}\left\{\Gamma(2N+1)\right\}^2\Gamma(s)}
          \int^{\infty}_0 d\lambda 
     \prod^{(n-3)/2}_{j=1/2}\left(\lambda^2+j^2\right)\frac{1}
       {{\rm e}^{2\pi\lambda}+1}\nonumber\\
      & &\times\int_{S_3} dz\:\cot(\pi z)
       \left(z^2-B_{\rm L}^2\right)^{-s+k} \,.    
        \label{eq;d-8}
\end{eqnarray}
Using the above definition, $\zeta_{\phi}(s)$ is then expressed as 
%
\begin{equation}
\zeta_{\phi}(s)=\tilde{\zeta}_2(s)+2(s-2)\:\tilde{\zeta}_1(s)
      -\hat{\zeta}_0(s)\,.
       \label{eq;d-9}
\end{equation}
 Now, we move a
contour $S_1$ to a parallel line along the imaginary axis, in order
to deal with the poles of $\cot(\pi z)$ in (\ref{eq;d-8})
 (see Fig.\,\ref{fig:d1}).
The contour $S_2$ is replaced with
 lines passing just above the cuts associated with $z=\pm A_{\rm L}$.
$\tilde{\zeta}(s)$ is then given by 
%
\begin{eqnarray}
\tilde{\zeta}_k(s)&=&
       \frac{\pi a^{2s}}{2^{2n-1}\left\{\Gamma(2N+1)\right\}^2\Gamma(s)}
       \left\{\left(\frac{a}{b_0}\right)^2
              {\rm e}^{-[2d/(n-2)+2]\kappa_\sigma \sigma}\right\}^{-s+k}
      \sin\left\{\pi\left(s-k\right)\right\}\nonumber\\
      & \times &\frac{1}{2(s-1)(s-2)}
      \left\{\int^{\infty}_0dx\;\:D_{\phi}(ix)
      \left(x^2+A_{\rm L}^2\right)^{-s+k}\coth(\pi x)\right.\nonumber\\
     & &\left.-\int^{A_{\rm L}}_0
      dx\:\cot(\pi x)\;D_{\phi}(x)\left(A_{\rm L}^2-x^2\right)^{-s+k}
      \right\}\,,
       \label{eq;d-10}
\end{eqnarray}
where $D_{\phi}(ix)$ is a polynomial with coefficients $r_{Nk}$:
%
\begin{eqnarray}
D_{\phi}(ix)&=&(-1)^{D}\frac{2x^2}{(2D)!}
        \left\{x^2+\left(D-1\right)^2\right\}\cdots
        \left\{x^2+1\right\}\nonumber\\
       &\equiv&(-1)^{D}\sum^{D-1}_{p=0}\;r_{Np}\;x^{2p+2}\,.
       \label{eq;d-11}
\end{eqnarray}
The first term in Eq.\,(\ref{eq;d-8}) comes from the integral 
along the imaginary
axis, and the second term in Eq.\,(\ref{eq;d-8}) 
is the contribution from the
contours along the cuts on the real axis.
Substituting the relation 
%
\begin{equation}
\coth(\pi x)=1+\frac{2}{{\rm e}^{2\pi x}-1}
       \label{eq;d-12}
\end{equation}
 into the first term of (\ref{eq;d-10}), 
the function $\tilde{\zeta}_k(s)$ is finally given by  
%
\begin{eqnarray}
\tilde{\zeta}_k(s)&
            =&\frac{\pi a^{2s}}{2^{2n-1}\left\{\Gamma(2N+1)\right\}^2
           \Gamma(s)}
          \left\{\left(\frac{a}{b_0}\right)^2
         {\rm e}^{-[2d/(n-2)+2]\kappa_\sigma \sigma}
\right\}^{-s+k}
        \sin\left\{\pi\left(s-k\right)\right\}\nonumber\\
      & &\times \frac{1}{2(s-1)(s-2)}     
      \left\{\frac{1}{2}\sum^{D-1}_{p=0}r_{Np}
      \left(A_{\rm L}^2\right)^{p+k+3/2-s}
       \frac{\Gamma\left(p+\frac{3}{2}\right)
      \Gamma\left(s-p-k-\frac{3}{2}\right)}
      {\Gamma\left(s-k\right)}\right.
      \nonumber\\
    & &-\int^{\infty}_0 dx\;D_{\phi}(ix)
       \left(x^2+A_{\rm L}^2\right)^{-s+k}
       \frac{2}{{\rm e}^{2\pi x}-1}\nonumber\\\
    & & \left.+\int^{A_{\rm L}}_0 dx D_{\phi}(x)
        \left(A_{\rm L}^2-x^2\right)^{-s+k}\cot(\pi x)\right\}\,.
         \label{eq;d-13}
\end{eqnarray}
Also, the contour $S_3$ is replaced with
 lines passing just above the cuts associated with $z=\pm B_{\rm L}$
 (see Fig.\,\ref{fig:d2}).
Using the procedure for calculating $\tilde{\zeta}_k(s)$, the function 
 $\hat{\zeta}_k(s)$ is then given by  
%
\begin{eqnarray}
\hat{\zeta}_k(s)&
            =&\frac{\pi a^{2s}}{2^{2n-1}\left\{\Gamma(2N+1)\right\}^2
           \Gamma(s)}
          \left\{\left(\frac{a}{b_0}\right)^2
         {\rm e}^{-[2d/(n-2)+2]\kappa_\sigma \sigma}\right\}^{-s+k}
        \sin\left\{\pi\left(s-k\right)\right\}\nonumber\\
      & &\times \int^{\infty}_0 d\lambda 
       \prod^{(n-3)/2}_{j=1/2}\left(\lambda^2+j^2\right)\frac{1}
       {{\rm e}^{2\pi\lambda}+1}\nonumber\\
     & &\times \left\{\frac{1}{2}\sum^{D-1}_{p=0}r_{Np}
      \left(B_{\rm L}^2\right)^{p+k+3/2-s}
       \frac{\Gamma\left(p+\frac{3}{2}\right)
      \Gamma\left(s-p-k-\frac{3}{2}\right)}
      {\Gamma\left(s-k\right)}\right.
      \nonumber\\
    & &-\int^{\infty}_0 dx\;D_{\phi}(ix)
       \left(x^2+B_{\rm L}^2\right)^{-s+k}
       \frac{2}{{\rm e}^{2\pi x}-1}\nonumber\\\
    & & \left.+\int^{B_{\rm L}}_0 dx D_{\phi}(x)
        \left(B_{\rm L}^2-x^2\right)^{-s+k}\cot(\pi x)\right\}\,.
         \label{eq;d-14}
\end{eqnarray}